\documentclass[aps,prc,twocolumn,groupedaddress]{revtex4-2}

\usepackage{graphicx}
\usepackage{slashed}
\usepackage{color}
\usepackage{amsmath}
\usepackage{multirow}
\usepackage{xspace}
\newcommand{\nopieft}{$\slashed{\pi}$EFT\xspace}

\newcommand{\be}{\begin{equation}}
\newcommand{\ee}{\end{equation}}
\newcommand{\nablavec}{\ensuremath{\boldsymbol{\nabla}}}
\newcommand{\LO}{\mathrm{LO}}

\newcommand{\bra}{\langle}
\newcommand{\ket}{\rangle}

\begin{document}
\title{Few nucleons scattering in pionless effective field theory}
\author{Martin Sch\"{a}fer}
\email{schafer.martin@mail.huji.ac.il}
\affiliation{The Racah Institute of Physics, The Hebrew University, 
               Jerusalem 9190401, Israel}
\author{Betzalel Bazak}
\email{betzalel.bazak@mail.huji.ac.il}
\affiliation{The Racah Institute of Physics, The Hebrew University, 
               Jerusalem 9190401, Israel}
\date{\today}

\begin{abstract}
We present a comprehensive theoretical study of low-energy few nucleon scattering for systems with $A\leq 4$.
To this end, we utilize pionless effective field theory, which we employ at next-to-leading order. We show that at this level the theory yields accurate predictions for the low-energy scattering parameters in all studied channels.
These predictions are on par with the best experimental evaluations and the available theoretical calculations. 
We confirm the recent observation that a four-body force is needed at next-to-leading-order and find that for nuclear systems it only appears in a single spin-isospin channel. 
\end{abstract}


\maketitle
\section{Introduction}
Effective field theories (EFTs) provide a thorough framework for the study of low-energy physics. 
In nuclear physics, EFTs are low-energy manifestations of the underlying theory, QCD.
As such, they are formulated in terms of baryons and mesons
as the fundamental degrees of freedom, rather than quarks and gluons, and are constructed to obey the symmetries of QCD. For a recent review see, e.g. \cite{HamKonKol20}.

Of particular interest is pionless EFT (\nopieft), which is the simplest possible nuclear EFT, having the mesons integrated out leaving the nucleons as the only degrees of freedom.
\nopieft is best suited to describe very low-energy processes 
and to explore universal physics, i.e. phenomena which are not sensitive to the details of the interparticle interaction and are therefore common to systems as different as nucleons and atoms.
As the deuteron binding energy is unnaturally small, light nuclei belong to a universality class where the scattering length is much larger than the interaction range. 
For a large and positive scattering length, a bound dimer exists whose energy is almost entirely determined by the scattering length alone. Besides the deuteron, another natural example of such a dimer is the He$_2$ molecule.

Behind its apparent simplicity, \nopieft exhibits some peculiar features, such as (a) The Wigner bound, which limits the possible values of the effective range and thus forces a perturbative treatment of all but leading-order (LO) terms \cite{Wigner}.
(b) The Thomas collapse, compelling the promotion of a 3-body contact term to LO \cite{Tho35, BedHamKol99}. 
(c) The Efimov effect \cite{Efi70}, dictating the 3-body states binding energy, and
(d) The appearance of a 4-body force at next-to-leading-order (NLO) \cite{BazKirKon19}.
Here, we would like to test the utility of \nopieft, comparing the predictions of the theory at NLO to experimentally measured low-energy 2-clusters $s$-wave scattering data of reactions involving $A\leq 4$ nucleons. To be more specific, 
as we use the $A=2$ nucleon-nucleon ($N+N$) scattering lengths and effective ranges,
as well as the triton binding energy, to fit the low-energy-constants (LECs) of the theory, we test \nopieft ability to predict the 3-body $N+2N$, and the 4-body $2N+2N$, and $N+3N$ low-energy scattering parameters. 

The application of \nopieft to study low-energy scattering dates back to its first days \cite{BedKol98,BedHamKol98}. However, so far, published works were predominantly limited to nuclear systems with $A\le 3$, i.e. nucleon-nucleon or nucleon-deuteron scattering. Here, we extend these studies considering perturbative insertion of next-to-leading order (NLO) terms and describing most $s$-wave scattering processes with $A\le 4$. 

When the scattering length is much larger than the interaction range, three 
identical bosons in an $s$-wave exhibit the Efimov effect.  
In the corresponding spin-half fermionic system, three particles in a relative $s$-wave are blocked due to the Pauli principle, and therefore Efimov effect appears only in higher partial waves \cite{Efi73,CasMorPri10,BazPet17}. 
Scattering of an atom from fermionic dimer exhibits, however, universal characteristic, and the atom-dimer scattering length $a_{a-dm}$ is determined by the atom-atom scattering length $a_{aa}$, $a_{a-dm} \approx 1.2 a_{aa}$ \cite{STM,GSS84,Pet03}. Moreover, the dimer-dimer scattering length is also universal and  $a_{dm-dm} \approx 0.6 a_{aa}$ \cite{PetSalShl04}.

In nuclear physics, where the nucleons carry two internal degrees of freedom,
spin and isospin, both phenomena are relevant. 
In some cases, 3 or 4 nucleons have symmetric spatial wave functions and therefore behave like a bosonic system, while in other cases they have antisymmetric one and thus behave like fermions, depending on the quantum numbers. To see that, let us consider three nucleons with zero total orbital angular momentum. 
In the spin-isospin $S,I=(1/2,1/2)$ channel the nucleons behave like bosons and therefore bound Efimovian three-body state exists and a three-body force is to be introduced to set its energy \cite{BedHamKol99}. In contrast, neutron-deuteron scattering in the $S,I=(3/2,1/2)$  channel is subject to Pauli blocking, and therefore universal fermionic behavior is expected \cite{STM}.

This is also the case for the four nucleon system. While deuteron-deuteron scattering in the $S,I=(2,0)$ channel has a fermionic nature, the $S,I=(0,0)$ channel is bosonic and a bound four-body state exists, i.e. $^4$He. This channel is of particular interest since it was shown recently that
for a bosonic system, a four-body force must be introduced at NLO to regularize the system \cite{BazKirKon19}. It is therefore interesting to check if such a force is needed also in the nuclear case, and if so in which channels.

Several studies have applied \nopieft to the $nd$ scattering \cite{BedKol98,BedHamKol98,BedGri00,GabBedGri00,HamMeh01,BedRupGri03,Gri04,Van13,MarSprVan16,Kon17,RupVagHig19}.
The spin-quartet $s$-wave channel was studied in Ref. \cite{BedKol98,BedHamKol98}, resuming effective-range corrections to all orders. The calculated $nd$ scattering length was found to be $6.33$ fm, in excellent agreement with the experimental value $6.35(2)$ fm \cite{DilKoeNis71}. 
The calculated phase shifts agreed with the phase shift analysis of Refs.~\cite{OerSea67,PhiBar69}. 
In the $S,I=(1/2,1/2)$ channel, a three-body parameter is needed and can be fitted, for example, to the triton binding energy.
Perturbative range corrections in this channel were considered in Ref.~\cite{HamMeh01}, and they were followed by calculations of higher-order contributions in \cite{BedRupGri03,Gri04}. A fully perturbative study of $nd$ scattering up to N$^2$LO was done in Ref.~\cite{Van13}, in a way that does not require the calculation of the full off-shell scattering amplitude, needed in earlier studies \cite{JiPhiPla12,JiPhi13}.
In the presence of Coulomb interaction, $pd$ scattering was studied as well \cite{KonHam14,VanEgoKer14,KonGreHam15}.

So far, four-nucleon scattering has not been explored extensively within the framework of \nopieft. The $p ^3$He and $nt$ scattering lengths were calculated at LO in Ref.~\cite{Kir13}. While the $p\,^3$He results were burdened by relatively large theoretical errors, the predicted $nt$ scattering lengths were found to be somewhat smaller than a compilation of phenomenological results presented there.  
A calculation of the spin-singlet $n^3$He scattering length was performed in Ref.~\cite{KirGriShu10} using a nonperturbative \nopieft NLO potential.  

Here, we present a systematic study of low-energy nuclear scattering up to $A \le 4$. We calculate scattering lengths and effective ranges for all available spin-isospin channels, to second order in \nopieft. 
To this end, we apply a shallow harmonic oscillator trap to the studied system, and extract the corresponding scattering parameters using the Busch formula \cite{BusEngRza98,SLB09}. Bound-state energies and wave functions with and without the trap are obtained employing a correlated Gaussian basis with the stochastic variational method (SVM) \cite{SuzVar98}.  

The paper starts with a short description of \nopieft in Sec.~II. In Sec.~III we present the numerical tools applied in our work. The fitting of \nopieft low-energy constants is briefly described in Sec.~IV. The results for different few-nucleons channels are presented in Sec.~V, followed by our conclusions in Sec.~VI.

\section{Model}

In this work, we study the few-nucleons scattering in the framework of \nopieft.
The dynamical degrees of freedom are nucleons, while pions, as well as other degrees of freedom, are integrated out, and the corresponding physics is encapsulated in the low energy constants. 

Since EFT contains all terms consistent with the symmetries of the underlying theory, a power counting, i.e. a scheme to determine which terms belong to each order of the theory, must be introduced in order to restore its predictive power. Naive power counting suggests only two-body $s$-wave contact interactions at LO, and therefore the relevant Lagrangian density should be
\be \begin{split}
\mathcal L_\LO = N^\dagger \left(i\partial_0 
+ \frac{\nablavec^2}{2m} \right) N
- &\frac{1}{2} C^{(0)}_0 (N^\dagger \hat{\mathcal P}^{0,1}_{2b} N)^2\\
- &\frac{1}{2} C^{(0)}_1 (N^\dagger \hat{\mathcal P}^{1,0}_{2b} N)^2. 
\end{split}
\ee
Here we set $\hbar=1$, $N$ is the nucleon field operator and $m$ its mass, $\hat{\mathcal P}^{S,I}_{2b}$ is a projection operator on a two-body channel with spin $S$ and isospin $I$, $C^{(0)}_0$ and $C^{(0)}_1$ are the low-energy constants.

In the three-body spin-isospin $S,I=(1/2,1/2)$ channel all nucleons can occupy the $s$-shell and the system collapses \cite{Tho35}. As a result, a three-body contact term must be introduced \cite{BedHamKol99},
\be 
\mathcal L^{(3b)}_\LO = - \frac{1}{6} D^{(0)} (N^\dagger \hat{\mathcal P}^{1/2,1/2}_{3b} N)^3 \,,
\ee
where $D^{(0)}$ is the three-body LEC and $\hat{\mathcal P}^{S,I}_{3b}$ is a projection operator into a tree-body spin-isospin channel.

At LO, the interaction is to be iterated, which is equivalent to solving the non-relativistic Schr\"odinger equation with the Hamiltonian 
\be \label{HLO}
H_\LO = -\frac{1}{2m}\sum_i \nabla^2_i + V_2^{(0)} + V_3^{(0)}\,,
\ee
where $V_2^{(0)}$ and $V_3^{(0)}$ are LO two- and three-body contact potentials, respectively.

The singular nature of contact interactions requires regularization, which is performed here by applying a local Gaussian regulator that suppresses momenta above an ultraviolet cutoff. 
Physical quantities predicted by the theory have to be independent of the cutoff since it is not physical quantity.  
This is achieved via renormalization, i.e. by fitting the values of the LECs to run with the cutoff in such a way that a chosen set of physical observables is reproduced. 
The regularized LO two-body potential then is
\be\label{v2}
V_2^{(0)} = \sum_{i<j} \left( C^{(0)}_0 \hat{\mathcal P}_{ij}^{0,1} + C^{(0)}_1 \hat{\mathcal P}_{ij}^{1,0} \right) g_\lambda(r_{ij})\,,
\ee
and the three-body potential is,
\be\label{v3}
V_3^{(0)} =  D^{(0)}_0 \sum_{i<j<k} \sum_{cyc} \hat{\mathcal P}^{1/2,1/2}_{ijk} g_\lambda(r_{ij})g_\lambda(r_{ik})\,,
\ee
where 
$g_\lambda(r) \propto \exp(-\lambda^2r^2/4)$ is the chosen regulator,
$cyc$ denotes the cyclic sum and $r_{ij} =|{\bf r}_i - {\bf r}_j|$ is a relative distance between nucleons $i$ and $j$. As $\lambda \rightarrow \infty$, the contact nature of the interaction is recovered.

The next-to-leading order contains range corrections, with new LECs $C^{(1)}_2$ and $C^{(1)}_3$ to be determined from two-body observables. 
The corresponding potential can be written as
\be \label{NLO} \begin{split}
& V^{(1)}_{2} = \\
& \sum_{i<j} \left( C^{(1)}_2 \hat{\mathcal P}_{ij}^{0,1} + C^{(1)}_3 \hat{\mathcal P}_{ij}^{1,0} \right)
 \Big( g_\lambda(r_{ij}) \overrightarrow{\nabla}^2_{ij} 
        + \overleftarrow{\nabla}^2_{ij}  g_\lambda(r_{ij})\Big)  \,.
\end{split} \ee

Unlike the LO interaction, which has to be treated non-perturbatively, NLO consists of a single insertion of the potential. Renormalization cannot be achieved for a positive effective range when an inconsistent subset of higher-order corrections is included by the nonperturbative solution of the Schr\"odinger equation with the NLO potential \cite{BeaCohPhi98}. 
Note that only recent studies of \nopieft scattering indeed treat NLO terms perturbatively.

Perturbative inclusion of range corrections, Eq.~\eqref{NLO}, changes renormalization conditions used at LO. As a result one has to consider perturbative inclusion of counter-terms in a form equivalent to Eqs.~\eqref{v2} and \eqref{v3}. This ensures that the renormalization conditions applied at LO remain satisfied at NLO and it leads to three other LECs -- $C_1^{(1)}$, $C_2^{(1)}$, and $D_0^{(1)}$. 

\begin{figure*}
\begin{tabular}{ccc}
\includegraphics[width=0.98\columnwidth]{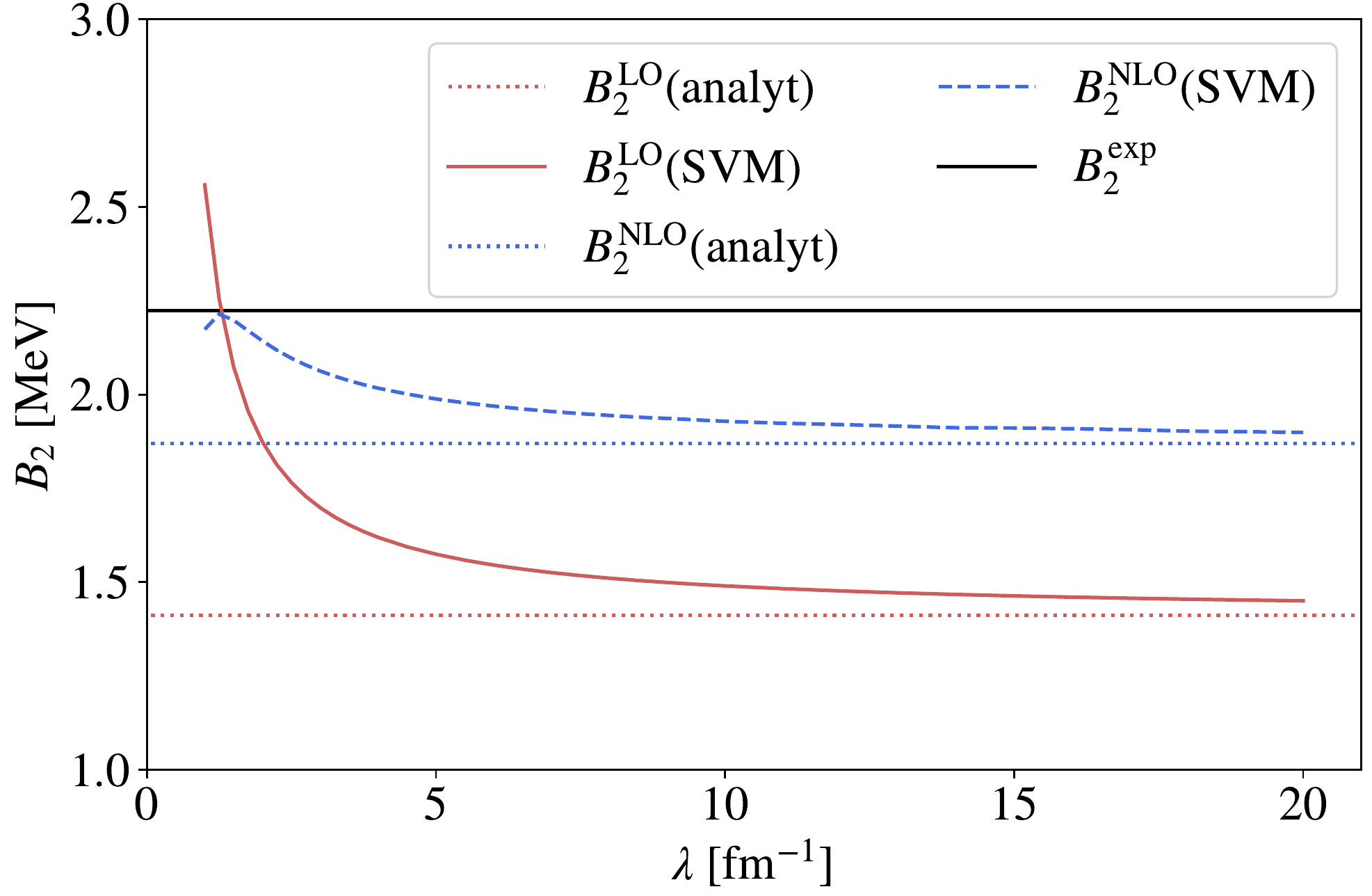}&~&\includegraphics[width=0.98\columnwidth]{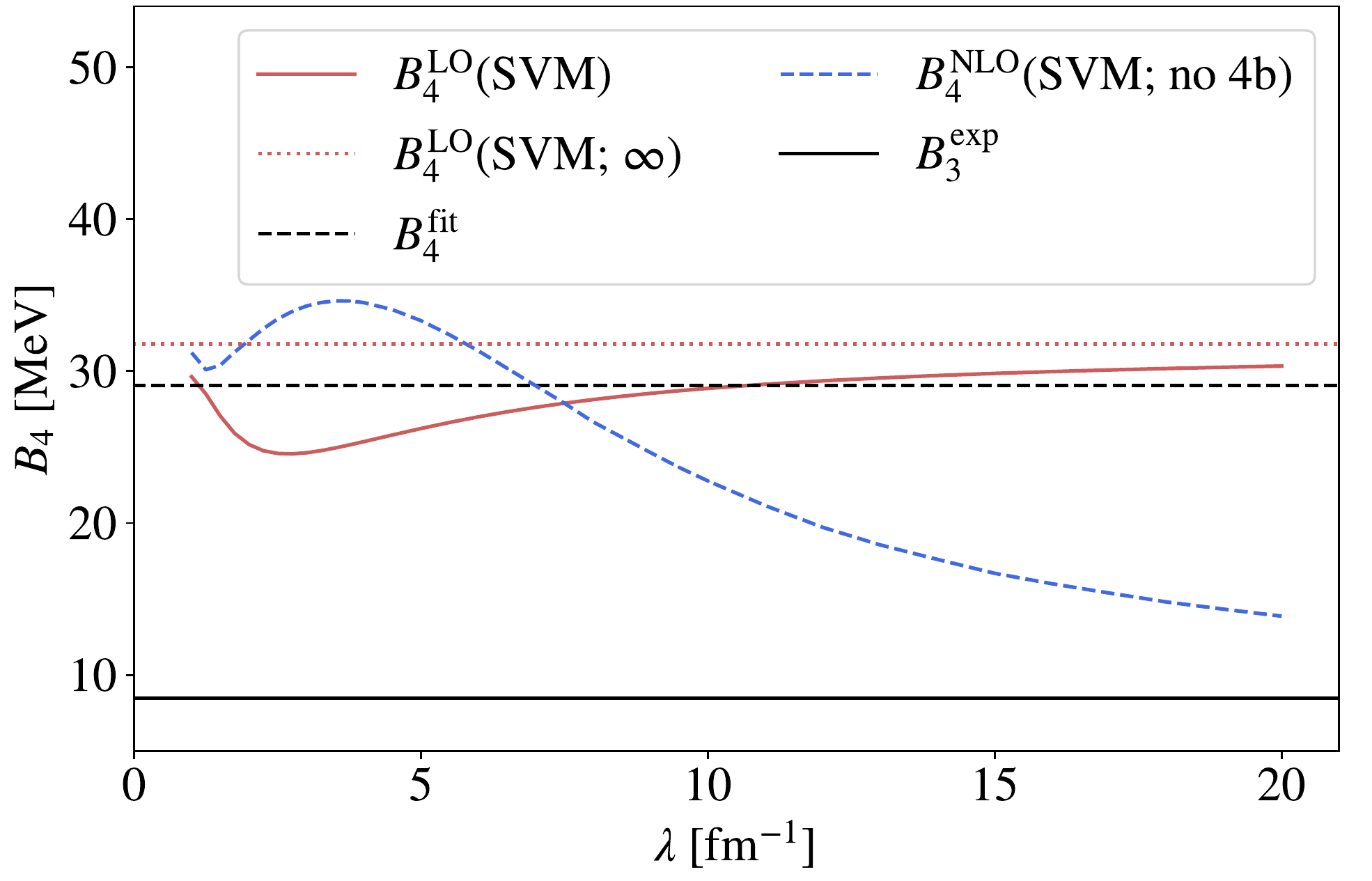}\\
\end{tabular}
\caption{\label{figHe4} Left panel: The deuteron binding energy as a function of the cutoff $\lambda$. The $B_2^{\rm LO}({\rm SVM})$ and $B_2^{\rm NLO}({\rm SVM})$ denote LO and NLO binding energies calculated using SVM. The $B_2^{\rm LO}({\rm analyt})$ and $B_2^{\rm NLO}({\rm analyt})$ values are predictions of Eq.~\eqref{b2h}. Right panel: The $^4$He binding energy as a function of the cutoff $\lambda$. The $B_4^{\rm LO}({\rm SVM})$ and $B_4^{\rm NLO}({\rm SVM;~no~4b})$ denote LO and NLO binding energies calculated using SVM with no four-body force considered at NLO. The $B_4^{\rm LO}({\rm SVM; \infty})$ stands for LO \nopieft binding energy extrapolated to $\lambda \rightarrow \infty$ using $B_{4}(\lambda)= B_{4}(\infty)+\alpha/\lambda$ function. The $B_4^{\rm fit}$ shows experimental $^4$He binding energy with Coulomb energy subtracted which is used to fix the NLO four-body force. The experimental triton binding energy $B_3^{\rm exp}$ is shown for comparison.}
\end{figure*}

For a bosonic case, it was shown recently that a contact four-body term has to be added at NLO \cite{BazKirKon19}. In this work, we demonstrate that indeed this is also the case for nuclei and a four-body term has to be considered in the $S,I=(0,0)$ four-body channel to obtain reasonable  results at this order. Here, we use hyperradial four-body force, 
\be\label{4b}
V_4 = E^{(1)}_0 \sum_{i<j<k<l} \hat{\mathcal P}^{0,0}_{ijkl}\, g_\lambda(r_{ijkl})\,,
\ee
where $r_{ijkl}$ is the four particles hyper radius,
$r_{ijkl}^2 = \sum_{\mu<\nu\in\{i,j,k,l\}}r^2_{\mu\nu}$
and $E^{(1)}_0$ denotes four-body LEC. Note, that the more economical hyperradial form of the four-body force was chosen over the one using a cyclic sum which includes many permutations.

\section{Method}
Few-body bound-state solutions are obtained by expanding the corresponding wave function $\Psi$ in a correlated Gaussian basis,
\be
  \Psi = \sum_{i} c_{i} ~\hat{\mathcal A}  \exp \left(-\frac{1}{2} {\bf x}^T \boldsymbol{\rm A}_i {\bf x}\right) \chi^i_{S M_S} \xi^i_{I M_I},
\ee
where the operator $\hat{\mathcal{A}}$ ensures antisymmetrization between nucleons, ${\bf x}^T=({\bf x}_1,\dots,{\bf x}_{A-1})$ is a set of Jacobi coordinates, and $\chi^i_{S M_S}$ and $\xi^i_{I M_I}$ stand for the relevant spin and isospin parts, respectively. Here $\boldsymbol{\rm A}_i$ is an $(A-1)\times(A-1)$ positive-definite symmetric matrix with $A(A-1)/2$ nonlinear parameters. 
Both bound-state energies and variational parameters $c_i$ are obtained by diagonalizing the Hamiltonian matrix. In order to choose basis states with the most appropriate nonlinear parameters, we use the SVM \cite{SuzVar98}, which was proved to offer an effective procedure to optimize the finite basis set, yielding a very accurate description of bound states.

Scattering states are not compact in space, which makes them more difficult to calculate. As a result, it might be easier to extract scattering parameters from bound-state calculations. To do so, one can apply periodic boundary conditions to the system at hand, calculate the discrete energies for a few box sizes, and then use the L\"uscher formulae to extract the scattering parameters in free space \cite{Luscher}.

A similar approach, which we use here, is to trap the studied nuclear $A$-body system in a harmonic oscillator (HO) potential 
\begin{equation}
    V_{\rm HO}(\textbf{r}) = \frac{m}{2A} \omega^2 \sum_{i<j} (\textbf{r}_i - \textbf{r}_j)^2\,,
\end{equation}
with oscillator frequency $\omega$.
Consider the scattering of two bound subclusters $B$ and $C$, and using a trap with length $b_{\rm HO} =\sqrt{2/(m \omega)}$ much larger than the other length scales of the system,
the subclusters then can be considered as point-like particles.
As a result, one can match the asymptotic part of the trapped wave function to 
the known solution of two trapped particles with a short-range interaction.
The $B C$ phase shifts $\delta_{\scriptstyle B C}$ at relative momentum $k$ then can be extracted using the Busch formula \cite{BusEngRza98,SLB09}  
\begin{equation}
    - \sqrt{4\mu \omega} \frac{\Gamma\left(3/4 - \epsilon^n_\omega/2\omega\right)}{\Gamma\left(1/4 - \epsilon^n_\omega/2\omega\right)} = k \cot \delta_{\scriptstyle B C} \,,
    \label{Busch}
\end{equation}
where $\mu = m_{B} m_{C} / (m_{B} + m_{C})$ is the $B C$ reduced mass, $\Gamma(x)$ is the Gamma function
, $k=\sqrt{2\mu \epsilon^n_\omega}$, and $\epsilon^n_\omega = E_\omega^n (A)-E_\omega (B)-E_\omega (C)$ is the $n$-th excited state energy of the trapped $A$-body system with respect to the $B + C$ threshold. Here, bound-state energies $E_\omega (B)$, $E_\omega (C)$, and $E_\omega^n (A)$ are calculated using the SVM. Throughout this work $BC \in \{NN, nd, nt, n^3{\rm He}, dd \}$ and all phase shifts are extracted applying HO trap lengths $15~{\rm fm} \leq b_{\rm HO} \leq 50$~fm. 

Note that one should consider energy levels above the trapped $B + C$ threshold disregarding possible $BC$ bound states. If there are any higher thresholds corresponding to further disintegration of $B$ or $C$ subclusters we select only such levels which correspond to $B C$ scattering in the HO trap.

The scattering length $a_{\scriptscriptstyle BC}$ and effective range $r_{\scriptscriptstyle BC}$ then result from fitting the calculated phase shifts with the effective range expansion (ERE)
\be \label{ere}
k \cot \delta_{\scriptstyle BC} = -\frac{1}{a_{\scriptscriptstyle BC}} + \frac{1}{2}r_{\scriptscriptstyle BC}k^2 + \ldots\;.
\ee
 
Next-to-leading order bound-state energies are obtained considering a single insertion of the NLO potential. Using first-order perturbation theory,
\be
E^{\rm NLO} = E^{\rm LO} + \frac{\left \bra \Psi^{\rm LO}~\right| V^{\rm NLO} \left| ~\Psi^{\rm LO} \right \ket}
{\left \bra \Psi^{\rm LO}~|~\Psi^{\rm LO}\right \ket}\,,
\label{pert}
\ee
where $V^{\rm NLO}$ stands for the sum of all NLO potential terms and $\Psi^{\rm LO}$ is the LO bound-state wave function. Scattering predictions at NLO are calculated by changing the trapped energies, using first-order perturbation theory, Eq.~\eqref{pert}, and then applying Eq.~\eqref{Busch} to extract the scattering parameters \cite{SRBK10,RSBK12}. This way one can avoid calculating the off-shell scattering matrix while still taking NLO in a perturbative way.

Since Eq.~\eqref{Busch} relies on the difference between energies in shallow traps, the corresponding values should be calculated with high precision. To do so we first use SVM to build several basis sets, each optimized for different trap lengths and cutoffs. 
These basis sets are then joined into a larger basis while omitting those states which are nearly linear dependent to maintain the numerical stability \cite{SchBazBar21}. Checking the convergence of the bound-state energies $E_\omega^n (A)$, $E_\omega (B)$, and $E_\omega (C)$, as well as NLO expectation values with an increasing amount of basis states, we verify that the overall accuracy is below $10^{-4}$~MeV. 

\section{Fitting the EFT}

Setting the numerical tools, we can fit the \nopieft LECs to reproduce relevant experimental observables.

The LO potential, Eq.~\eqref{HLO}, has 2 two-body LECs, $C^{(0)}_0$ and $C^{(0)}_1$, corresponding to 2 $s$-wave $NN$ channels, spin-singlet ($^1S_0$) and spin-triplet ($^3S_1$). 
We fit these LECs to reproduce the experimental spin-singlet neutron-neutron scattering length $a^0_{nn}=-18.95$~fm \cite{nnSCTlngth1,nnSCTlngth2} and spin-triplet proton-neutron scattering length $a^1_{pn}=5.419$~fm \cite{machleidt01}, respectively. Since LO terms are iterated to all orders, one can apply standard tools to calculate the scattering length; we use the variable phase method \cite{VPM} as well as extracted it from phase shifts calculated with the Numerov method. 

At NLO, each two-body channel has a new momentum-dependent term, Eq.~\eqref{NLO}, thus the effective ranges can be reproduced as well. Since NLO terms have to be treated perturbatively, we utilize the distorted-wave Born approximation (DWBA) to fit the LECs. To preserve leading order renormalization conditions, we consider a perturbative correction to the LO interaction. This results in $2$ LECs in the $NN$ $^1S_0$ channel, $C^{(1)}_0$ and $C^{(1)}_2$, and $2$ LECs in the $NN$ $^3S_1$ channel, $C^{(1)}_1$ and $C^{(1)}_3$, fitted to reproduce experimental neutron-neutron spin-singlet effective range $r^0_{nn}=2.75$~fm \cite{MNS90} and proton-neutron spin-triplet effective range $r^1_{pn}=1.753$~fm \cite{machleidt01}, respectively, while keeping the same values of scattering lengths as have been fitted at LO. 

\begin{figure*}
\begin{tabular}{ccc}
\includegraphics[height=0.93\columnwidth]{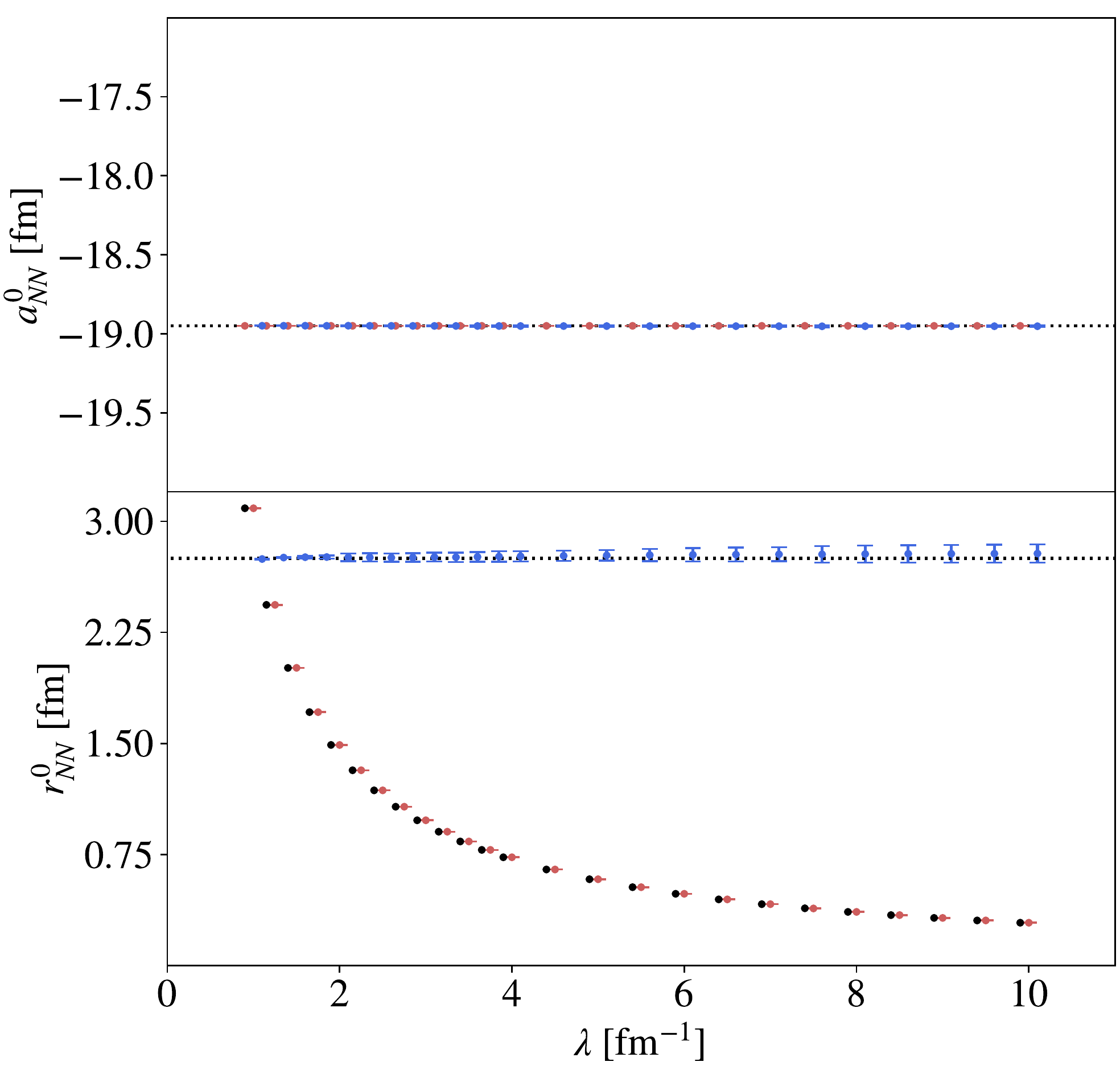}&~\hspace{5pt}~&\includegraphics[height=0.93\columnwidth]{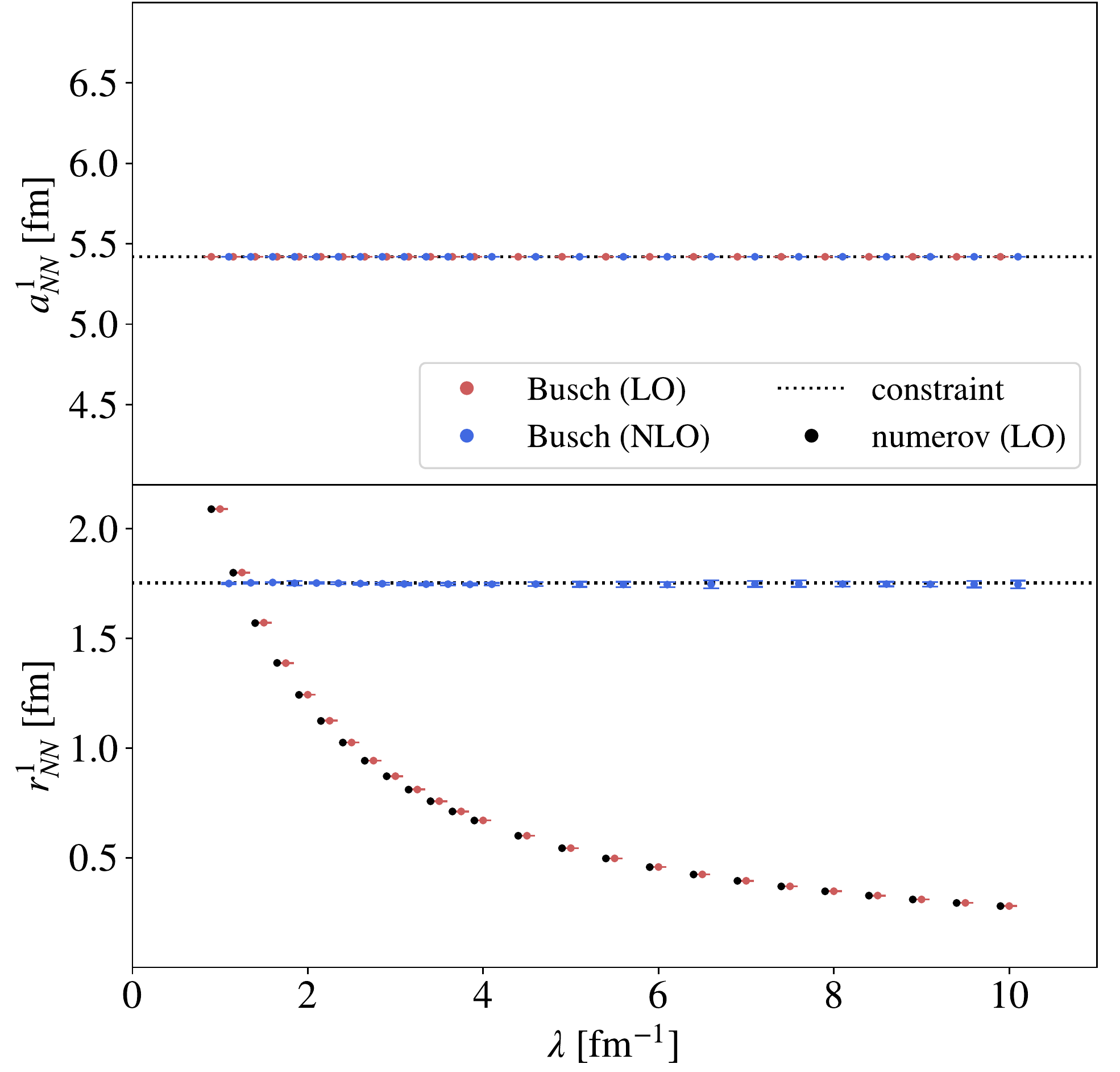}\\
\end{tabular}
\caption{\label{figNN}Left panel: Spin-singlet $NN$ scattering length $a^0_{NN}$ and effective range $r^0_{NN}$ values extracted from LO (red) and NLO (blue) phase shifts calculated via Busch formula, Eq.~\eqref{Busch}, as a function of increasing momentum cutoff $\lambda$. Dotted lines show experimental $a^0_{NN}$ and $r^0_{NN}$ constraints used to fix LO and NLO LECs. Black dots stand for LO $r^0_{NN}$ calculated via the Numerov algorithm. All LO Busch results are slightly shifted to the right in order to avoid complete overlap by displaying Busch NLO $a^0_{NN}$ and Numerov LO $r^0_{NN}$ results. Right panel: The same as in the left panel but for spin-triplet $NN$ channel.}
\end{figure*}

As we fit the two-body potential to scattering data, the deuteron binding energy $B_2$ is a prediction of the EFT. In the left panel of Fig.~\ref{figHe4} we show LO and NLO $B_2$ values calculated via SVM as a function of the cutoff. 
For zero-range attractive interaction, the LO and NLO deuteron binding energy can be calculated analytically
\be 
B_2^{\rm LO}=\frac{1}{m (a^1_{NN})^2},~~~~B_2^{\rm NLO}=B_2^{\rm LO}\left(1+\frac{r^1_{NN}}{a^1_{NN}}\right)\,.
\label{b2h}
\ee
Corresponding binding energy values are shown as red (LO) and blue (NLO) dotted lines in the left panel of Fig.~\ref{figHe4}. Clearly, our SVM results converge to these values as $\lambda\longrightarrow\infty$.

To constrain the LO three-body term we use SVM to solve for the triton binding energy and fit its $D^{(0)}_0$ LEC to reproduce the experimental value, $B_3=8.482$~MeV \cite{tritonBE}. Although no new three-body term appears at NLO, the LO term has to be corrected by introducing a three-body counter-term. Using the LO triton wave function, the corresponding LEC $D^{(1)}_0$ is fitted perturbatively keeping the NLO triton binding energy at its experimental value, while inserting the two-body NLO interaction. 

At LO, the $^4$He binding energy $B_4$ is correlated to that of the triton, and thus no four-body force is needed. However, it was shown recently \cite{BazKirKon19} that this is not the case at NLO for a bosonic system, where a contact four-body term has to be added to regularize the four boson system. Here, we show that this is also the case for nuclei. The right panel of Fig.~\ref{figHe4} shows $B_4$ as a function of the cutoff. The LO results, calculated via SVM, indeed converge at large $\lambda$s to a reasonable value. However, the perturbative inclusion of NLO corrections without the 4-body force gives much smaller binding energy. One should notice that for $\lambda \ge 10~{\rm fm}^{-1}$ such NLO corrections induce change larger than 30\% with respect to the LO result which casts certain doubts on the validity of a perturbative approach. This reveals the necessity of a four-body force, Eq.~\eqref{4b}, at NLO. Its LEC $E^{(1)}_0$ is perturbatively fitted, using SVM LO $^4$He wave function to get $B_4=29.046$ MeV \cite{heliumBE}, corresponding to the experimental value with Coulombic energy subtracted. 

\begin{figure*}
\begin{tabular}{ccc}
\includegraphics[height=0.93\columnwidth]{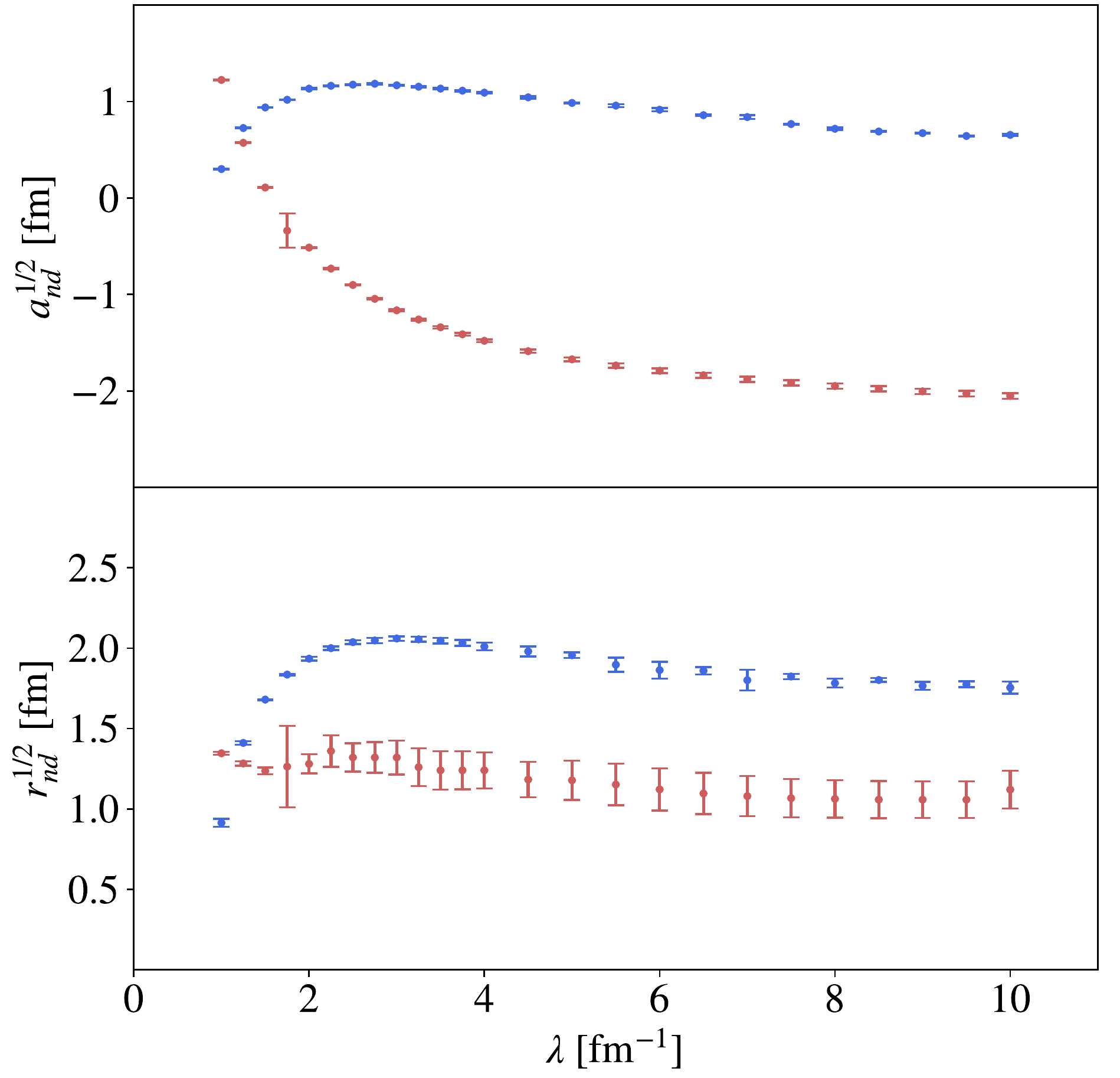}&~\hspace{5pt}~&\includegraphics[height=0.93\columnwidth]{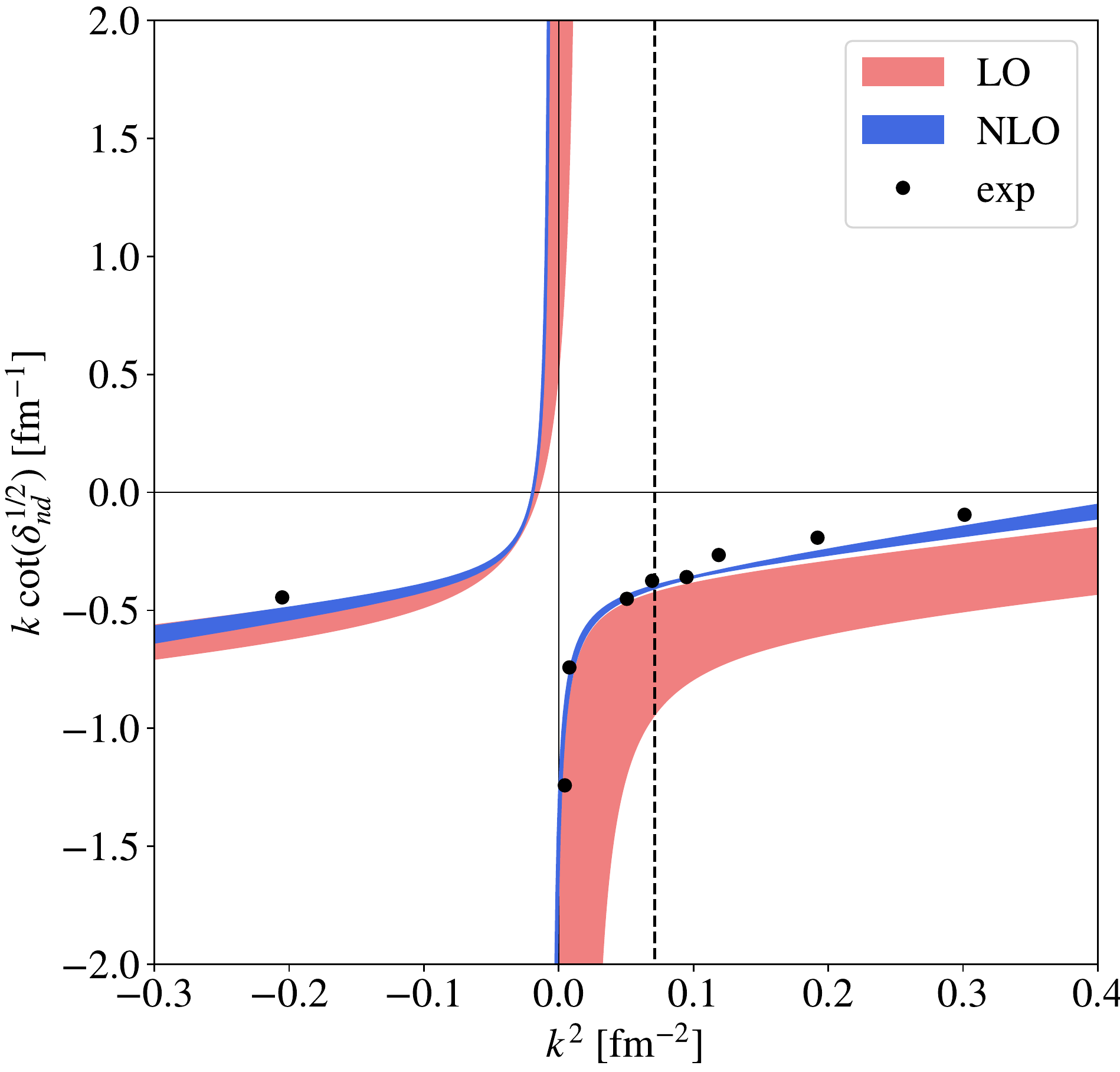}\\
\end{tabular}
\caption{\label{figN2N12}
Left panel: Spin-doublet $nd$ scattering lengths $a^{1/2}_{nd}$ and effective ranges $r^{1/2}_{nd}$ at LO (red) and NLO (blue), extracted using modified ERE, Eq.~\eqref{mere}, as a function of increasing momentum cutoff $\lambda$. Right panel: Possible LO (red) and NLO (blue) $k\cot\delta_{nd}^{1/2}$ values suggested by cutoff variation for $\lambda \in [1.25,10]~{\rm fm^{-1}}$. Black dots represent experimental phase shifts from Ref.~\cite{OerSea67}, the $k^2<0$ point relates to the triton bound state.
A dashed black vertical line marks the deuteron breakup threshold.
}
\end{figure*}

\section{Results}

Equipped with the tools needed to study few-nucleon elastic scattering, we performed accurate calculations for $A \leq 4$ nuclear systems within LO and NLO \nopieft. Considering a wide range of momentum cutoffs $\lambda \in [1,10]~{\rm fm^{-1}}$ we assess the cutoff dependence of our results. Our predictions are then thoroughly compared to the available experimental data.

\subsection{Two nucleons}

To check our numerical procedure, we consider two-nucleon $s$-wave scattering in the $^1S_0$ and $^3S_1$ channels. As mentioned above, corresponding experimental values of scattering lengths and effective ranges are used as an input to fit LO and NLO two-body LECs, respectively. 
We calculate here the low-energy scattering parameters from the phase shift extracted from SVM energies in a harmonic trap using the Busch formula.

In Fig.~\ref{figNN} we show the $NN$ spin-singlet (left panel) and spin-triplet (right panel) scattering lengths and effective ranges calculated using the Busch formula as a function of the momentum cutoff $\lambda$. For LO \nopieft potential, we obtain cutoff independent scattering length values in agreement with experimental constraints used in the LO fit. As we approach the limit of zero-range interaction $\lambda \rightarrow \infty$, the effective ranges $r^0_{NN}$ and $r^1_{NN}$ converge to zero. Direct comparison between their residual values calculated at finite $\lambda \in [1,10]~{\rm fm^{-1}}$ using the Busch formula and those extracted using the Numerov algorithm reveals basically identical results. 

Including perturbative NLO terms into the Busch formula yields the same values of scattering lengths as at LO. Calculated effective ranges, $r^0_{NN}$ and $r^1_{NN}$, are at NLO cutoff independent and in agreement with corresponding experimental values which have been used to fix $C_0^{(1)}$, $C_1^{(1)}$, $C_2^{(1)}$, and $C_3^{(1)}$ LECs via the distorted-wave Born approximation.

\subsection{Three nucleons}
For a system of three nucleons, there are two spin-isospin channels $S,I=(1/2,1/2)$ (doublet) and $S,I=(3/2,1/2)$ (quartet) which describe $s$-wave elastic scattering between deuteron bound state and the remaining nucleon. 
In all other channels, there is no two-body bound state.

\subsubsection{Doublet channel ($S=1/2$)}

In the spin-doublet channel, all nucleons are allowed to occupy the $s$-shell, and thus a three-body repulsive force is needed to prevent the Thomas collapse \cite{Tho35, BedHamKol99}. 

A characteristic feature here is the presence of a node in the $S$-matrix located close to the threshold. This leads to an anomalous behavior of corresponding scattering phase shifts: once displayed in terms of $k \cot \delta_{nd}^{1/2}$ as a function of $nd$ relative momentum squared $k^2$ they are dominated by the presence of a pole and conventional ERE does not apply. Hence, one must consider the modified effective range expansion \cite{OerSea67}
\be
k \cot \delta_{nd}^{1/2} = -A + \frac{B}{2} k^2 - \frac{C}{1 + D k^2} + \dots\,,
\label{mere}
\ee
where $a_{nd}^{1/2}=1/(A+C)$, $r_{nd}^{1/2}=B$, 
and the last term accounts for a pole position at $k^2 = -1/D$.

\begin{figure}
\includegraphics[width=8.6 cm]{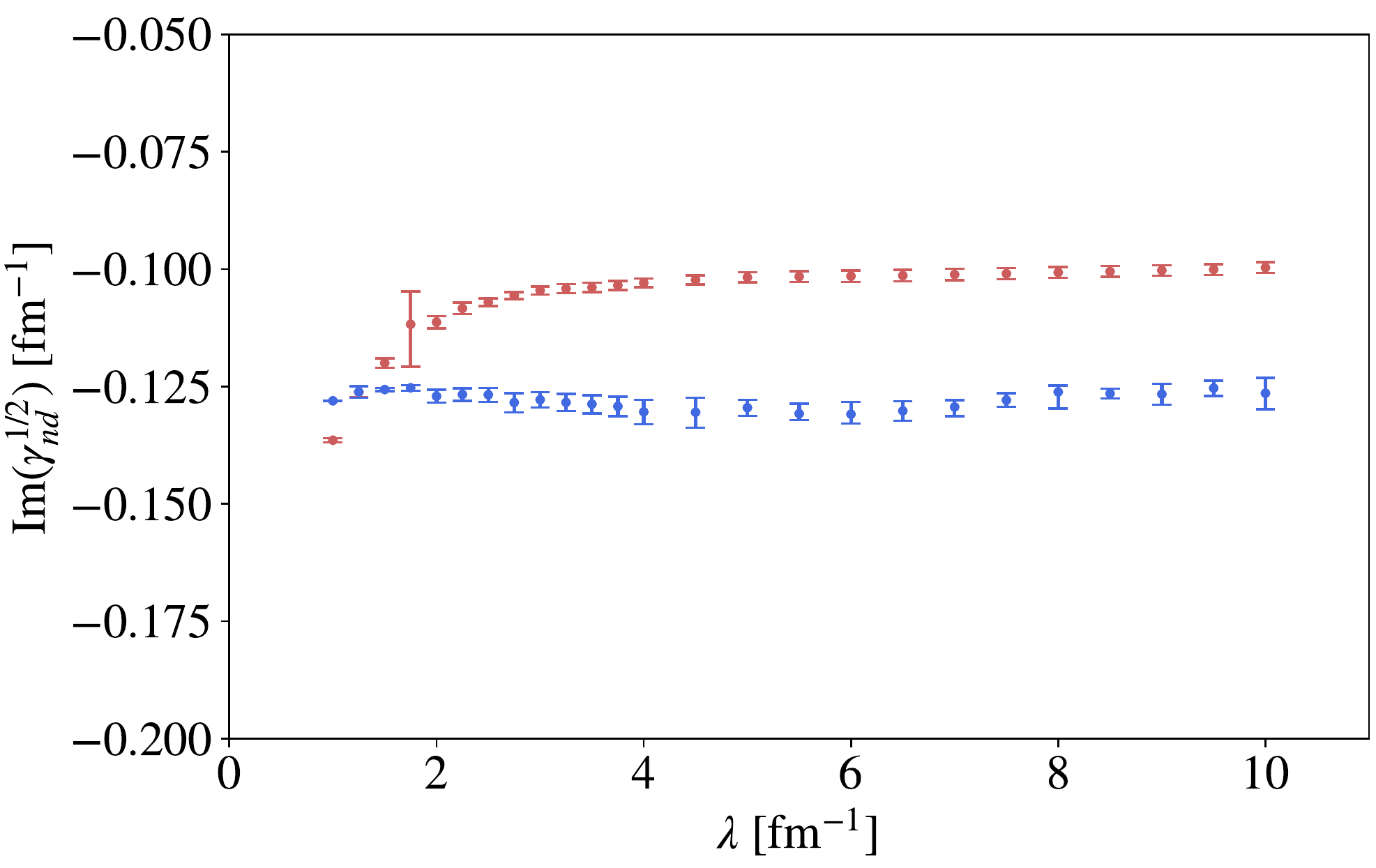}
\caption{\label{fig3Hvirt} Imaginary part of the triton virtual state momentum ${\rm Im}(\gamma_{nd}^{1/2})$ as a function of momentum cutoff $\lambda$ obtained as a root of $k\cot\delta_{nd}^{1/2}-{\rm i} k=0$. Here, $k\cot\delta_{nd}^{1/2}$ is represented by modified ERE, Eq.~\eqref{mere}, and fitted to calculated LO (red) and NLO (blue) $nd$ phase shifts in $S=1/2$ $s$-wave channel.}
\end{figure}

\begin{figure*}
\begin{tabular}{ccc}
\includegraphics[height=0.93\columnwidth]{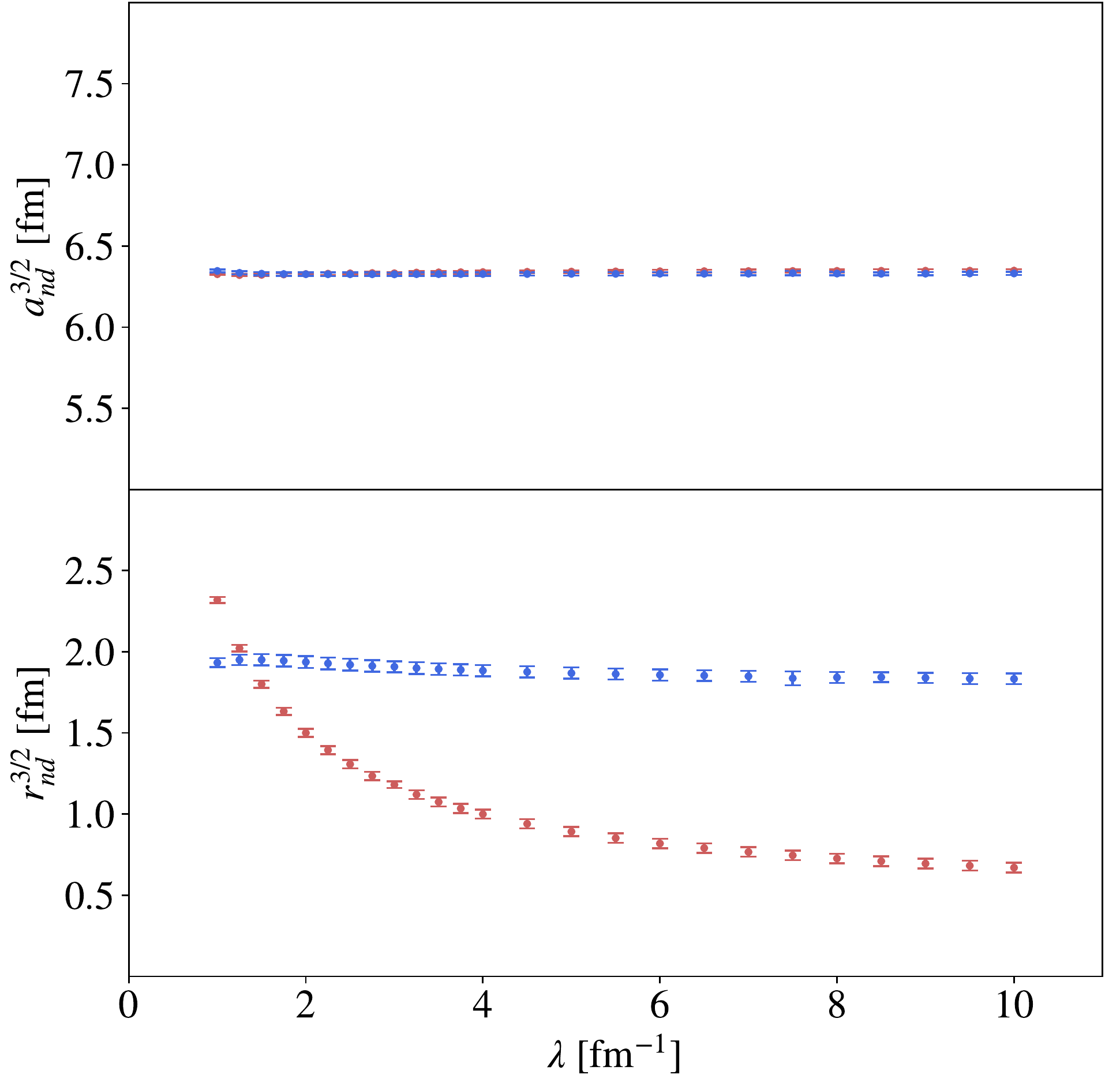}&~\hspace{5pt}~&\includegraphics[height=0.93\columnwidth]{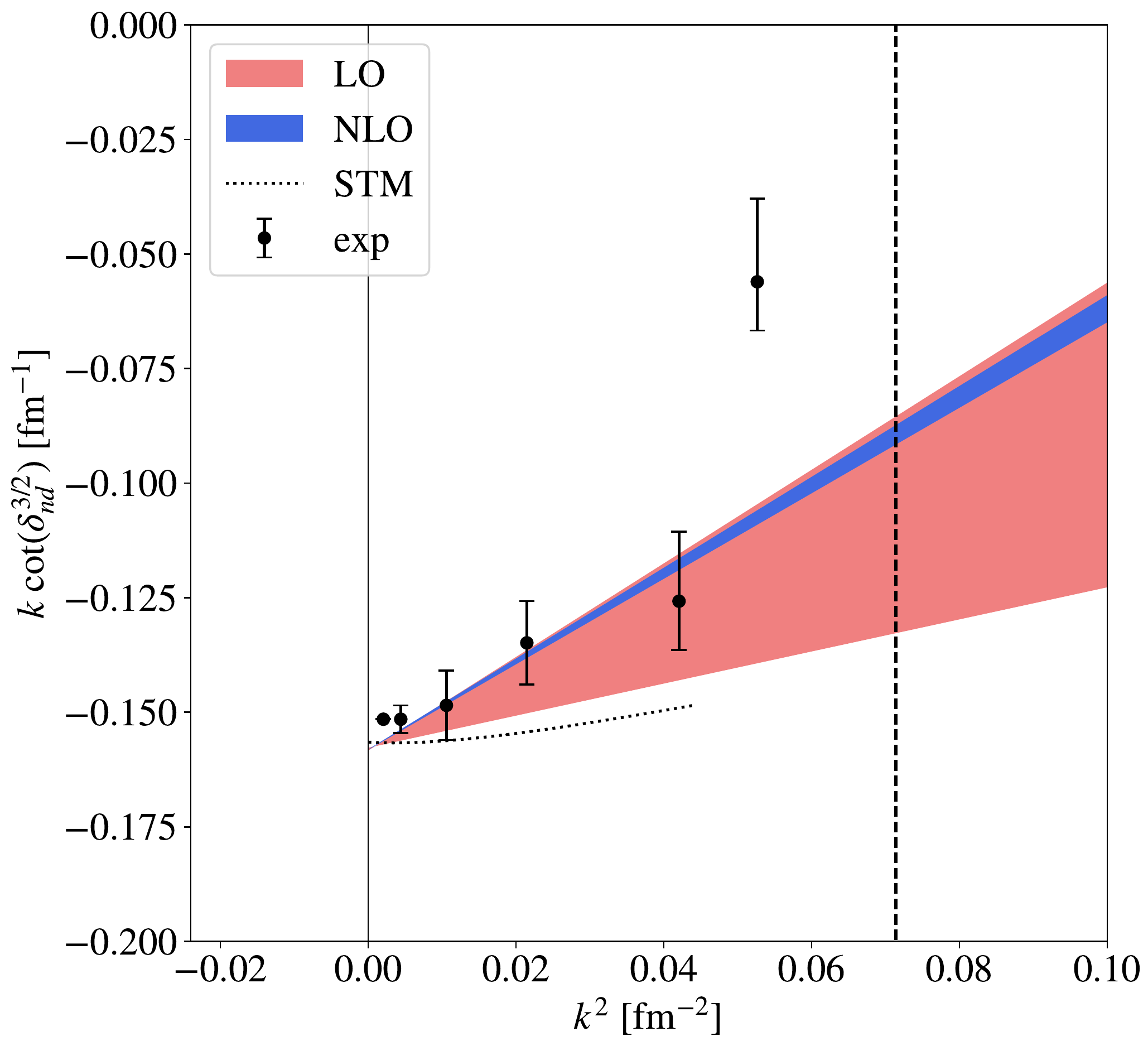}\\
\end{tabular}
\caption{\label{figN2N34}The same as Fig.~\ref{figN2N12}, but for spin-quartet $nd$ scattering. Black error bars in the right panel represent experimental phase shifts from Ref.~\cite{PhiBar69}. The dotted black line in the same panel shows scattering prediction obtained by solving the STM equation \cite{STM}.}
\end{figure*}

\begin{figure}
\includegraphics[width=8.6 cm]{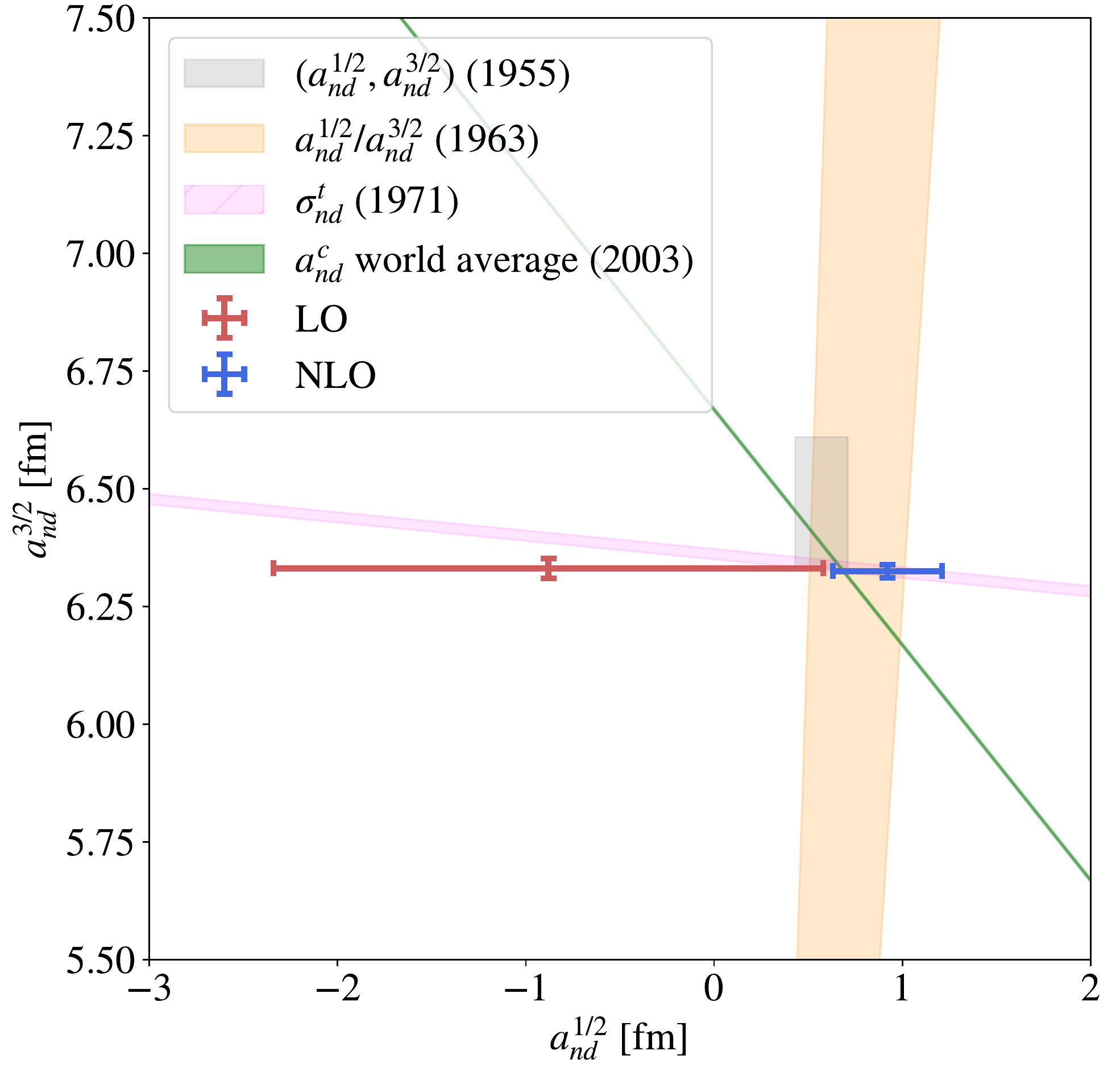}
\caption{\label{figN2NALL}
The $(a^{1/2}_{nd}; a^{3/2}_{nd})$ plane with experimental constraints imposed by different measurements -- size of $a^{1/2}_{nd}$ and $a^{3/2}_{nd}$ (grey) \cite{nikitin55}, $a^{1/2}_{nd}/a^{3/2}_{nd}$ ratio (yellow) \cite{gissler63}, $nd$ zero-momentum total cross-section (pink) \cite{PBS80}, and world average of experimental coherent scattering length (green) \cite{schoen03}. Our \nopieft LO and NLO predictions are shown by red and blue error bars, respectively.}
\end{figure}

We calculate the spin-doublet $nd$ scattering phase shifts at LO and NLO for different momentum cutoffs $\lambda$. Inspection of the corresponding $k\cot\delta_{nd}^{1/2}$ values reveals considerable dependence on the $nd$ relative momentum $k^2$ in agreement with the pole structure assumed in Eq.~\eqref{mere}. 
In order to explore the effect of the pole on the $nd$ scattering, we fit calculated LO and NLO $k\cot\delta_{nd}^{1/2}$ values using Eq.~\eqref{mere}. 
At LO, the pole position moves quite rapidly with an increasing cutoff from the subthreshold region $k^2<0$ above the $nd$ threshold $k^2>0$ and starts to converge at $\lambda \geq 4~{\rm fm^{-1}}$. The inclusion of NLO corrections stabilizes the pole position just below the $nd$ threshold. 
In the left panel of Fig.~\ref{figN2N12} we show possible $k \cot \delta_{nd}^{1/2}$ values given by a residual cutoff dependence for $\lambda \in [1.25,10]~{\rm fm^{-1}}$. One can see that once NLO corrections are included, a good agreement of our results with the experimental data \cite{OerSea67} is achieved.

The LO and NLO $a_{nd}^{1/2}$ and $r_{nd}^{1/2}$ values, extracted from the fit via Eq.~\eqref{mere}, are shown as a function of the cutoff in the left panel of Fig.~\ref{figN2N12}. Larger uncertainties in LO $r_{nd}^{1/2}$ results are induced by the dominance of the pole term in the modified ERE in a small momentum region between the $nd$ threshold and the deuteron breakup threshold. Considering residual cutoff variation for $\lambda \in [1.25,10]~{\rm fm^{-1}}$ and the numerical accuracy of our results, we obtain
\be
\begin{split}
    \text{LO}~~  &:~~ a^{1/2}_{nd}=-0.9(1.5)~{\rm fm}~~~~~r^{1/2}_{nd}=1.20(26)~{\rm fm},\\[5pt]
    \text{NLO} &:~~ a^{1/2}_{nd}=0.92(29)~~{\rm fm}~~~~~~r^{1/2}_{nd}=1.74(33)~{\rm fm}.\\
\end{split}
\nonumber
\ee

Having access to the fitted parameters of Eq.~\eqref{mere}, we search for roots of the equation $k\cot\delta_{nd}^{1/2}-{\rm i}k=0$. This provides us with a momentum $\gamma_{nd}^{1/2}$ of the first triton excited state which we find in a form of near-threshold virtual state (${\rm Re}(\gamma_{nd}^{1/2})=0;~ {\rm Im}(\gamma_{nd}^{1/2})<0$). In Fig.~\ref{fig3Hvirt} we show ${\rm Im}(\gamma_{nd}^{1/2})$ results at LO and NLO as a function of $\lambda$. Including NLO corrections largely suppresses a residual cutoff dependence. Taking into account numerical accuracy and $\lambda$ variation we obtain at LO ${\rm Im}(\gamma_{nd}^{1/2})=0.117(19)~{\rm fm^{-1}}$ and at NLO ${\rm Im}(\gamma_{nd}^{1/2})=0.1271(39)~{\rm fm^{-1}}$, corresponding to energies of  $E_{nd}^{1/2}=-0.43(14)~{\rm MeV}$ and $E_{nd}^{1/2}=-0.503(31)~{\rm MeV}$, respectively. Our results agree rather well with LO \nopieft prediction $E_{nd}^{1/2}=-0.574~{\rm MeV}$ of Ref.~\cite{GVHK19} and energy predicted by a separable potential model $E_{nd}^{1/2}=-0.48~{\rm MeV}$ \cite{AFT82}.

\subsubsection{Quartet channel ($S=3/2$)}
In the spin-quartet channel, two nucleons are allowed to occupy the $s$-shell forming $S=1$ deuteron bound state. The third nucleon is Pauli blocked preventing Thomas collapse and thus no three-body repulsive force is necessary. Moreover, LO and NLO potentials, Eqs.~(\ref{v2}),(\ref{NLO}), contribute to this channel only through $s$-wave spin-triplet two-body interaction which provides us with a resemblance to a fermionic atom-dimer scattering.

We calculate the scattering phase shifts in the spin-quartet $nd$ channel at LO and NLO. Since there is no anomalous behavior such as in the doublet case, corresponding scattering length and effective range, $a^{3/2}_{nd}$ and $r^{3/2}_{nd}$, are extracted using standard ERE, Eq. (\ref{ere}). In the left panel of Fig.~\ref{figN2N34} we show LO and NLO $a^{3/2}_{nd}$, $r^{3/2}_{nd}$ values as a function of momentum cutoff $\lambda$. Considering again residual cutoff variation and numerical accuracy of our predictions we get
\be
\begin{split}
    \text{LO}~~  &:~~ a^{3/2}_{nd}=6.336(21)~{\rm fm}~~~~~r^{3/2}_{nd}=1.43(79)~{\rm fm},\\[5pt]
    \text{NLO} &:~~ a^{3/2}_{nd}=6.330(14)~{\rm fm}~~~~~r^{3/2}_{nd}=1.891(91)~{\rm fm}.\\
\end{split}
\nonumber
\ee

Negligible effect of NLO corrections on $a^{3/2}_{nd}$ can be understood by the universal relations for the atom-dimer scattering length $a_{a-dm}$ which is determined by the atom-atom scattering parameters $a_{aa}$ and $r_{aa}$ \cite{STM,GSS84,Pet03},
\be
\frac{a_{a-dm}}{a_{aa}} = 1.179066 - 0.03595 \frac{r_{aa}}{a_{aa}}
\label{aduni}
\ee
Using the scattering parameters of the spin-triplet $NN$ channel this corresponds to $a^{3/2}_{nd}=6.389$~fm at LO and $a^{3/2}_{nd}=6.326$~fm at NLO, in a good agreement with our results. 

The universal relation for the atom-dimer effective range is \cite{STM,GSS84,Pet03}
\be
\frac{r_{a-dm}}{a_{aa}} = -0.0383 + 1.0558 \frac{r_{aa}}{a_{aa}},
\ee
which yields at LO 
$r^{3/2}_{nd}=-0.207$~fm and  
$r^{3/2}_{nd}=1.643$~fm at NLO.
Extrapolating our LO $r^{3/2}_{nd}$ results to the zero range $\lambda \rightarrow \infty$ limit yields $r^{3/2}_{nd}(\infty)=0.451(1)$~fm, in agreement with the results of Ref.~\cite{SteGreBlu08}, but apparently different from the universal results.

To better understand this issue we solve the Skorniakov-Ter-Martirosian (STM) equation \cite{STM} to obtain LO $nd$ spin-quartet phase shifts in the zero-range limit.
The corresponding solution is depicted using a black dotted curve in the right panel of Fig.~\ref{figN2N34}. 
The negative effective range predicted by Eq.~\eqref{aduni} can be seen only in quite small scattering energies; the shape parameters, giving the next corrections in the effective range expansion, bend the curve upwards for larger energies, which might lead to the positive effective range predicted by finite-range potentials. 

\subsubsection{Zero-momentum $nd$ scattering}
So far our predictions of the $s$-wave $nd$ elastic scattering have been compared to scarce experimental results of Ref.~\cite{OerSea67}. In order to access experimental information for zero-momentum, $a^{1/2}_{nd}$ and $a^{3/2}_{nd}$ in particular, several experiments were performed which provide certain constraints on the possible scattering length values $(a^{1/2}_{nd}; a^{3/2}_{nd})$.

In Fig.~\ref{figN2NALL} we show the $(a^{1/2}_{nd}; a^{3/2}_{nd})$ plane with the experimental constraints imposed by the following measurements: size of $a^{1/2}_{nd}$ and $a^{3/2}_{nd}$ (grey) \cite{nikitin55}, $a^{1/2}_{nd}/a^{3/2}_{nd}$ ratio (yellow) \cite{gissler63}, experimental value of $nd$ zero-momentum total cross-section (pink) \cite{DilKoeNis71}
\be
\sigma^t_{nd} = 4\pi \left[ \frac{1}{3}\left(a_{nd}^{1/2}\right)^2+\frac{2}{3}\left(a_{nd}^{3/2}\right)^2\right]
\ee
and the world average of experimental $nd$ coherent scattering length (green) \cite{schoen03}
\be
a_{nd}^c = \frac{1}{3}a_{nd}^{1/2}+\frac{2}{3}a_{nd}^{3/2}.
\ee
Our LO and NLO predictions of $(a^{1/2}_{nd}; a^{3/2}_{nd})$ scattering lengths are shown using red and blue error bars, respectively. Corresponding uncertainties are induced by numerics and residual $\lambda \in [1.25,10]~{\rm fm^{-1}}$ dependence of our results. 

Our $nd$ scattering length predictions can be further compared to the values extracted from experimental data. The modified ERE, Eq.~\eqref{mere}, was used to fit experimental spin-doublet $nd$ phase shifts extracting scattering length $a^{1/2}_{nd}=0.29~{\rm fm}$ \cite{OerSea67}. Somewhat larger experimental scattering length $a^{1/2}_{nd}=0.65(4)~{\rm fm}$ was extracted from zero-energy scattering measurements together with its spin-quartet equivalent $a^{3/2}_{nd}=6.35(2)~{\rm fm}$ \cite{DilKoeNis71}. Using a compilation of $a^{3/2}_{nd}$ theoretical predictions more precise $a^{1/2}_{nd}=0.645 \pm 0.003({\rm exp.}) \pm 0.007({\rm theory})~{\rm fm}$ value was obtained from the world average coherent scattering length \cite{schoen03}.   

\subsection{Four nucleons}
A system of four nucleons might exist in 9 different spin-isospin channels where both the total spin $S$ and the total isospin $I$ can get any value from $0,~1,$ and $2$. Among these channels only $S,I=(0,0)$ allows all four nucleons to occupy the $s$-shell and it is solely this channel that contributes, within our framework, to the $\rm ^4 He(0^+)$ ground state. 
As shown earlier, the inclusion of two- and three-body NLO corrections to $B_4$ leads to results outside of validity of perturbation theory and which deviate from the experimental value. Following Ref.~\cite{BazKirKon19}, we assume that NLO cutoff invariance in a nuclear case might be recovered by a perturbative inclusion of 4-body force. However, it is not clear whether the remaining $S,I$ four-body channels are affected as well or the necessity of 4-body force remains pertinent only to the $S,I=(0,0)$ channel with none of the nucleons being Pauli blocked.

In order to explore cutoff dependence we study $s$-wave elastic $3+1$ scattering in $S,I = (0,0)$, $(1,0)$, $(0,1)$, and $(1,1)$ four-body channels and $2+2$ scattering in $S,I = (2,0)$ four-body channel at LO and NLO \nopieft. Here, $2N$ and $3N$ stand for the deuteron and triton ground state, respectively. Remaining channels $S,T = (0,2)$, $(1,2)$, $(2,1)$, and $(2,2)$ are not considered in this work since they do not support partition into two subclusters.

\subsubsection{S=2 channel}

\begin{figure}
\includegraphics[width=8.6 cm]{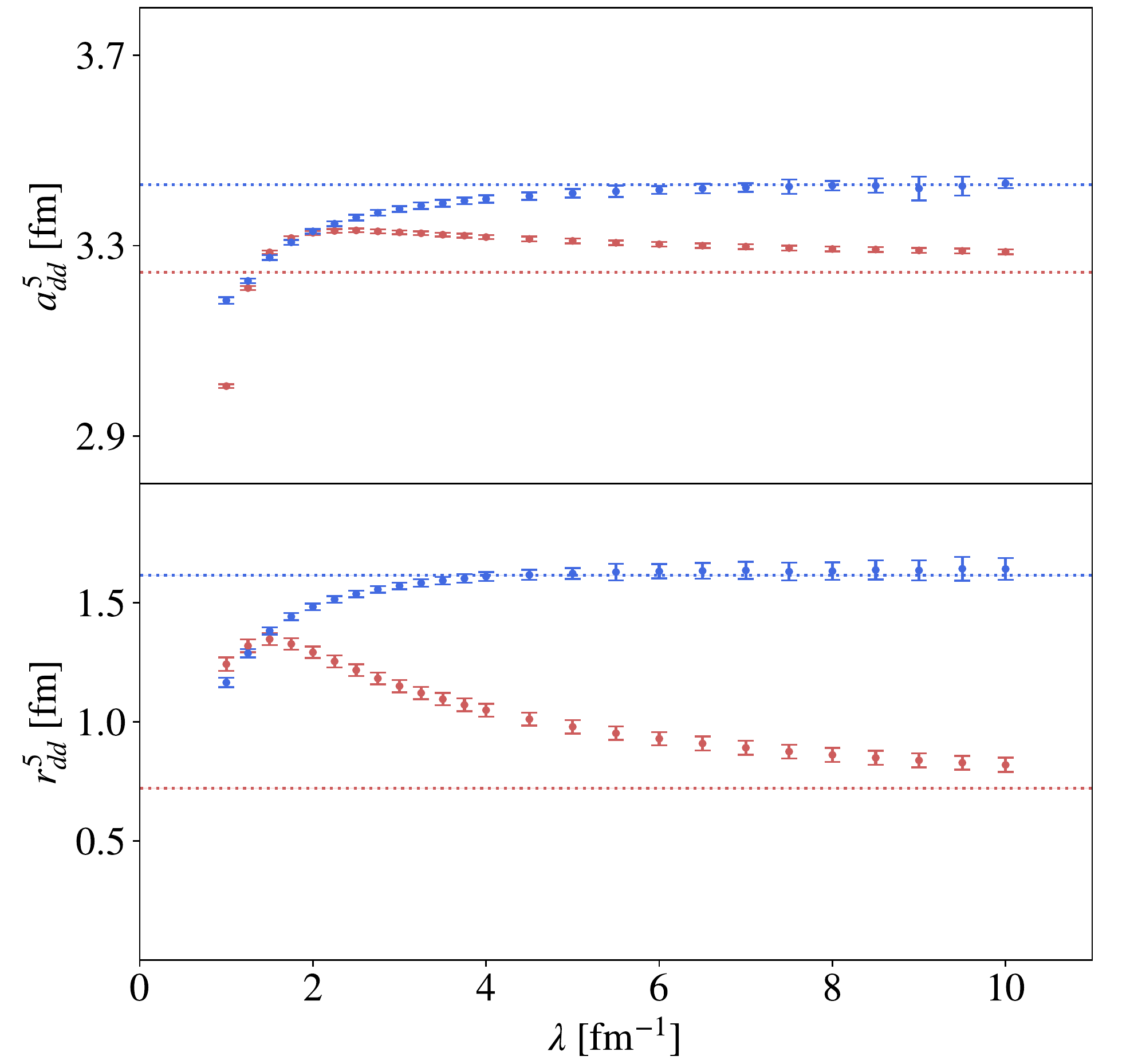}
\caption{\label{fig5S2}
The $dd$ scattering length $a^5_{dd}$ and effective range $r^5_{dd}$ in spin-isospin $S,I=(2,0)$ four-body channel at LO (red) and NLO (blue) as a function of the momentum cutoff $\lambda$. Error bars denote the numerical uncertainty of our results. Red and blue dotted lines marks predicted dimer-dimer scattering parameters using Eq.~\eqref{deltuva} for $a_{aa},r_{aa} = (5.419,0)$~fm and $a_{aa},r_{aa} = (5.419,1.753)$~fm, respectively.}
\end{figure}

The $S,I=(2,0)$ channel allows four-body partition into two deuteron bound states. Here, we disregard Coulomb interaction, consequently, we do not compare our results of $S=2$ $s$-wave ($^5S_2$) $dd$ elastic scattering to Refs.~\cite{Car21,FilYak00} where Coulomb force enters strongly at low momenta. 
Instead, similarly as for $S=3/2$ $nd$ scattering, at LO and NLO \nopieft only $s$-wave $NN$ spin-triplet interaction contributes and our calculations can be analyzed through universality in fermionic dimer-dimer scattering. 
The corresponding fermionic dimer-dimer scattering length $a_{dm-dm}$ and effective range $r_{dm-dm}$ can be parameterized through the atom-atom scattering parameters $a_{aa}$ and $r_{aa}$ by ~\cite{PetSalShl04,Del17}
\be \label{deltuva}
\begin{split}
\frac{a_{dm-dm}}{a_{aa}} &= 0.5986 + 0.105~\frac{r_{aa}}{a_{aa}} \pm 0.0005;\\
\frac{r_{dm-dm}}{a_{aa}} &= 0.133 + 0.51~\frac{r_{aa}}{a_{aa}} \pm 0.002.
\end{split}
\ee
Here, we repeat the procedure already used to analyze spin-quartet $S=3/2$ $nd$ scattering. For LO (zero range) $NN$ spin-triplet scattering parameters Eq.~\eqref{deltuva} yields $a_{dm-dm}=3.2438(5)$~fm and $r_{dm-dm}=0.721(2)$~fm and for the NLO case $a_{dm-dm}=3.4278(5)$~fm and $r_{dm-dm}=1.615(2)$~fm. If $^5S_2$ $dd$ elastic scattering exists in a universality window, i.e. range of validity of Eq.~\eqref{deltuva}, we expect that calculated LO and NLO deuteron-deuteron scattering parameters should converge for $\lambda \rightarrow \infty$ to these values of $a_{dm-dm}$ and $r_{dm-dm}$.      

\begin{figure*}
\begin{tabular}{ccc}
\includegraphics[height=0.63\columnwidth]{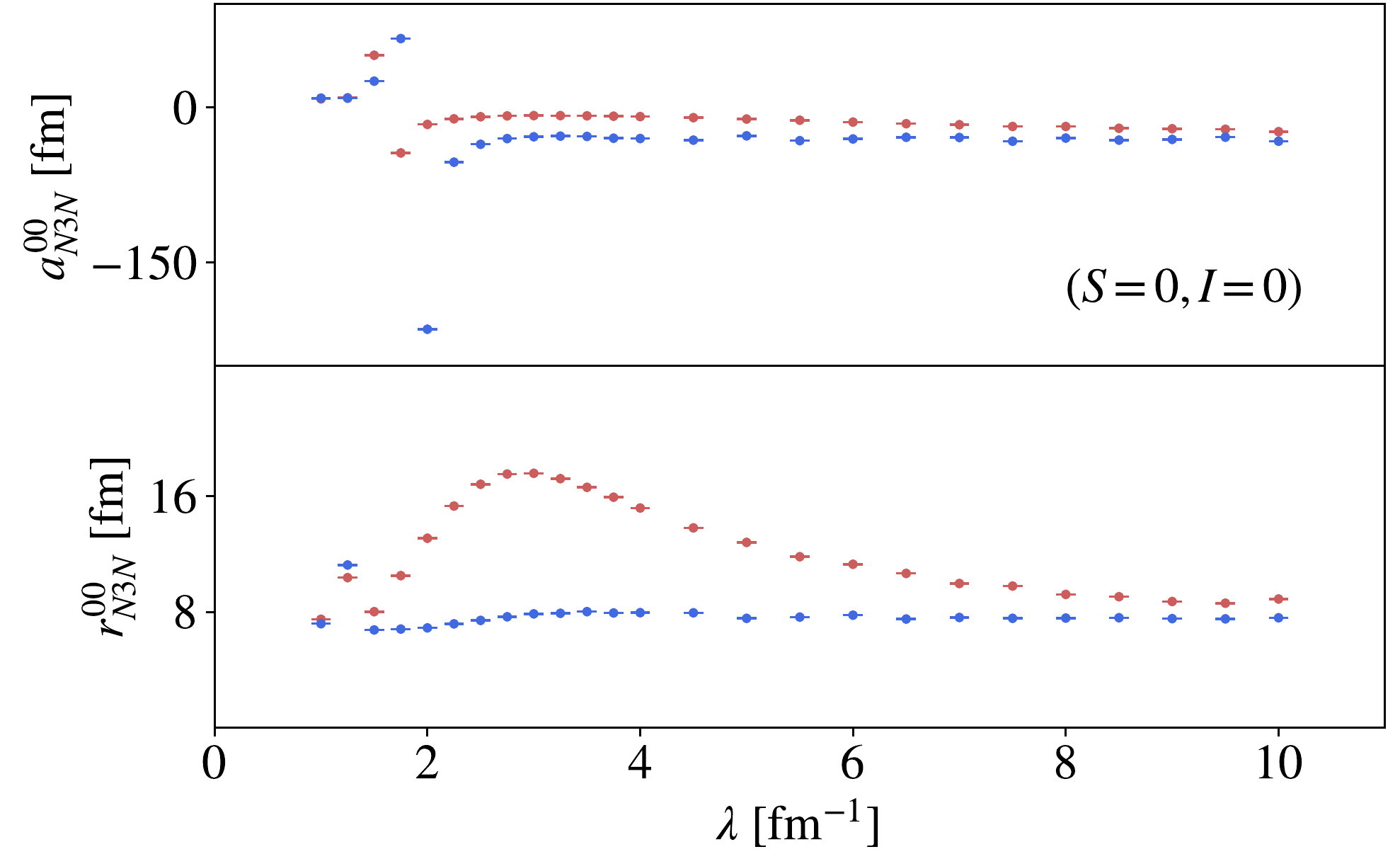}~~&~~&\includegraphics[height=0.63\columnwidth]{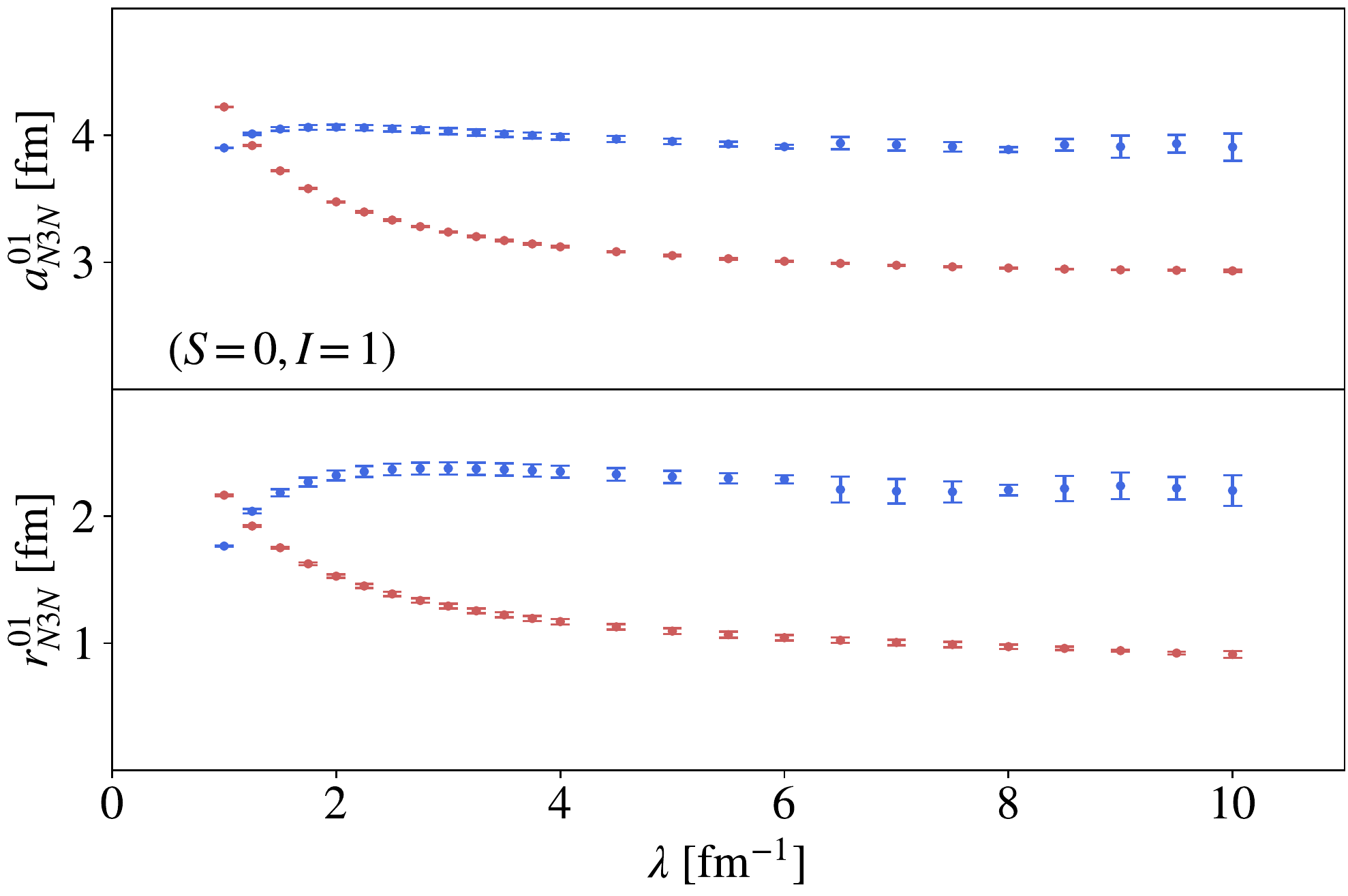}\\
\includegraphics[height=0.63\columnwidth]{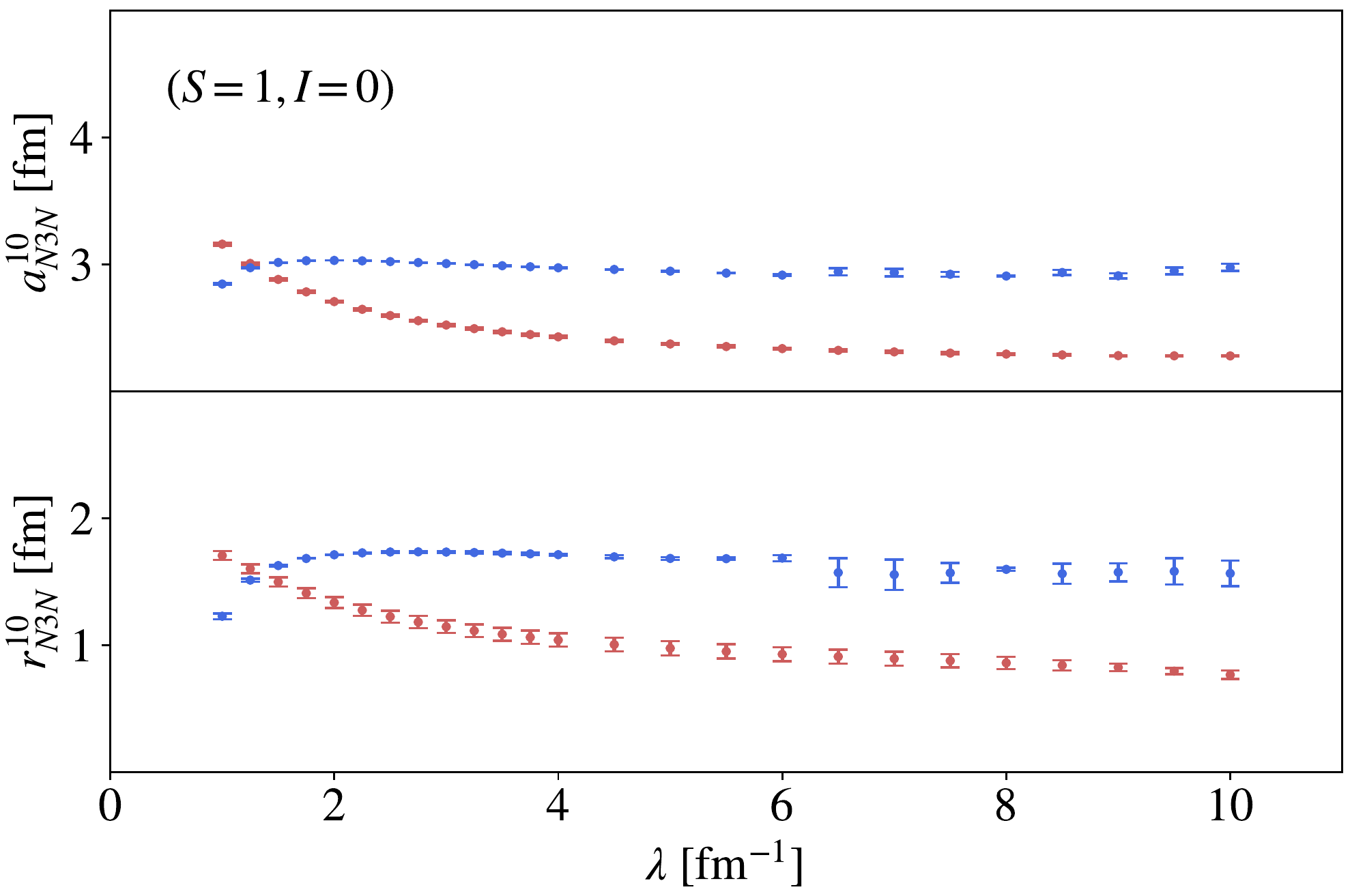}&~~&\includegraphics[height=0.63\columnwidth]{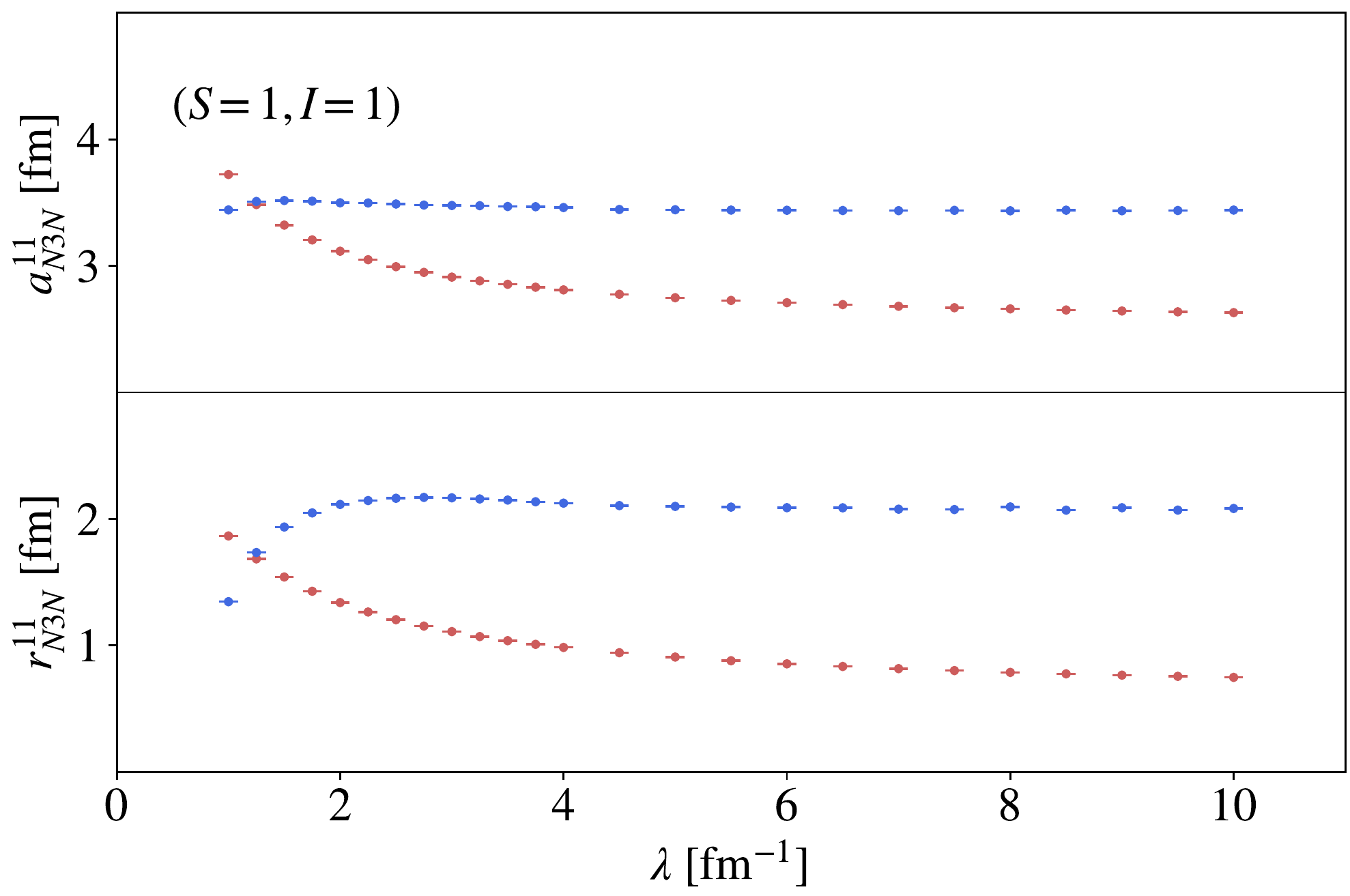}\\
\end{tabular}
\caption{\label{figN3Nar}
$N3N$ scattering lengths $a_{N3N}^{SI}$ and effective ranges $r_{N3N}^{SI}$ at LO (red) and NLO (blue), as a function of increasing momentum cutoff $\lambda$. Scattering parameters are shown for $S,I = (0,1)$, $(1,1)$, $(1,0)$, and $(0,0)$ spin-isospin four-body channels. Error bars denote the numerical uncertainty of our results.}
\end{figure*}

We calculate $dd$ phase shifts in $^5S_2$ channel at LO and NLO \nopieft. Corresponding scattering lengths $a_{dd}$ and effective ranges $r_{dd}$ are shown as a function of $\lambda \in [1,10]~{\rm fm^{-1}}$ in Fig.~\ref{fig5S2}. We observe that both LO and NLO predictions converge with increasing $\lambda$ to $a_{dm-dm}$ and $r_{dm-dm}$ values predicted via Eq.~\eqref{deltuva}. Our calculations thus demonstrate that within \nopieft, elastic $dd$ scattering in $^5S_2$ channel is in line with the universal prediction of fermionic dimer-dimer scattering of Refs.~\cite{PetSalShl04,Del17}, moreover, our NLO results show no requirement of 4-body force in this channel.

\subsubsection{S=0 and S=1 channel}
Moving to $3+1$ scattering, we first check cutoff dependence in $S,I = (0,0)$, $(1,0)$, $(0,1)$, and $(1,1)$ four-body channels. In Fig.~\ref{figN3Nar} we show the scattering lengths and effective ranges extracted from the calculated $s$-wave $N-3N$ phase shifts. We find that in $S,I = (1,0)$, $(0,1)$, and $(1,1)$ four-body channels there is no requirement of additional 4-body force. In $S,I=(0,0)$ channel the 4-body force, Eq.~\eqref{4b}, is included and, as can be seen from the upper left panel of Fig~\ref{figN3Nar}, our NLO results indeed converge with $\lambda$. 

With no Coulomb interaction, our $N-3N$ results in Fig.~\ref{figN3Nar} might be regarded as $n t$ or $n{\rm ^3He}$ scattering. Adopting full isospin symmetry the $n t$ system might scatter either in $S,I=(0,1)$ (spin-singlet) or $S,I=(1,1)$ (spin-triplet) four-body channel. 
For $n{\rm ^3He}$, the corresponding $s$-wave spin-singlet and spin-triplet scattering has contribution from two four-body spin-isospin channels $S,I=(0,1)+(0,0)$ and $S,I=(1,1)+(1,0)$, respectively. To the best of our knowledge, there are no available $s$-wave low momentum experimental phase shifts below the $t$/$^3$He breakup threshold. Consequently, only the zero momentum part of our results, in terms of scattering lengths, will be compared to a collection of available experimental data and theoretical works.

\begin{figure*}
\begin{tabular}{ccc}
\includegraphics[height=0.93\columnwidth]{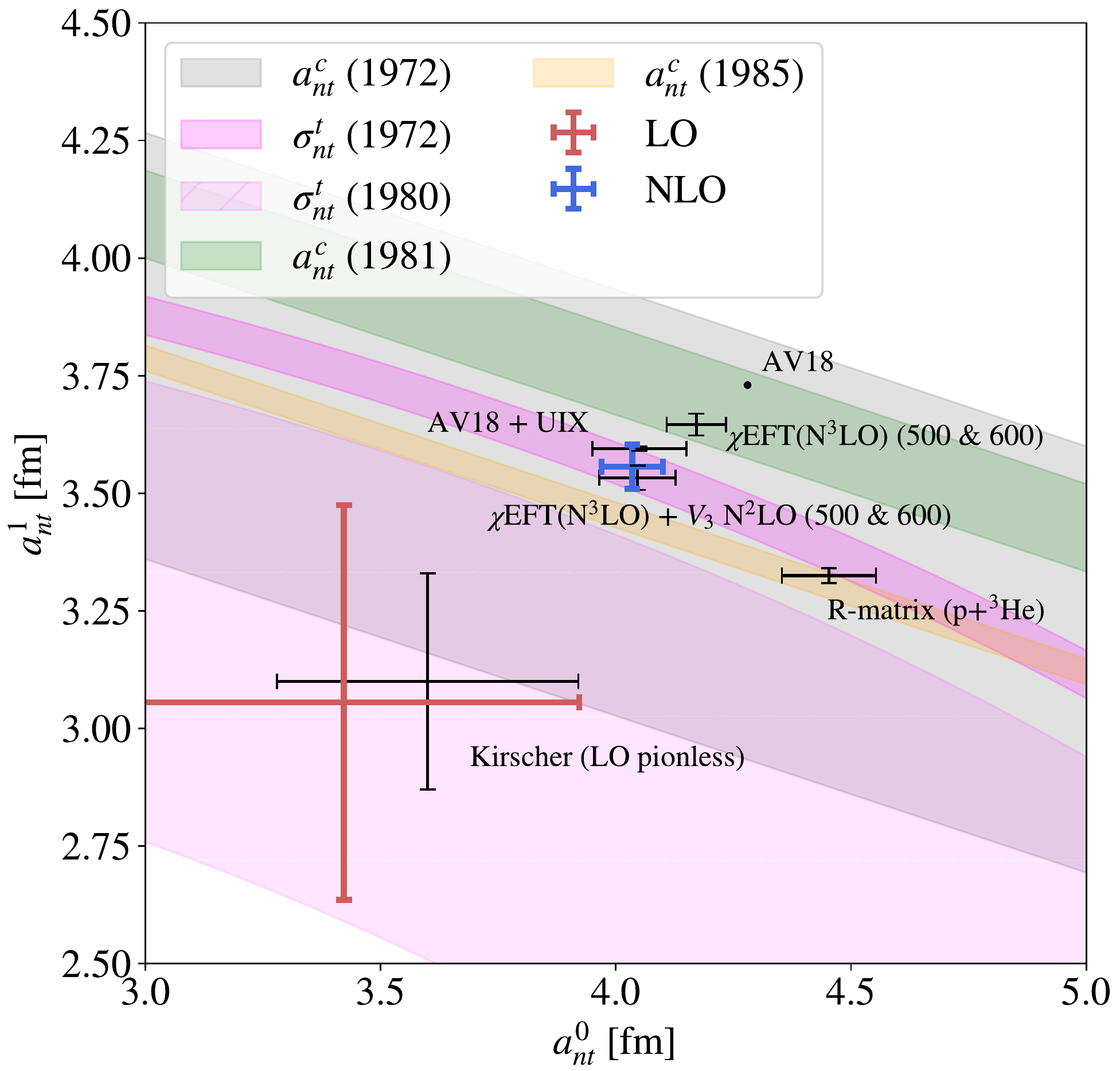}&~\hspace{5pt}~&\includegraphics[height=0.93\columnwidth]{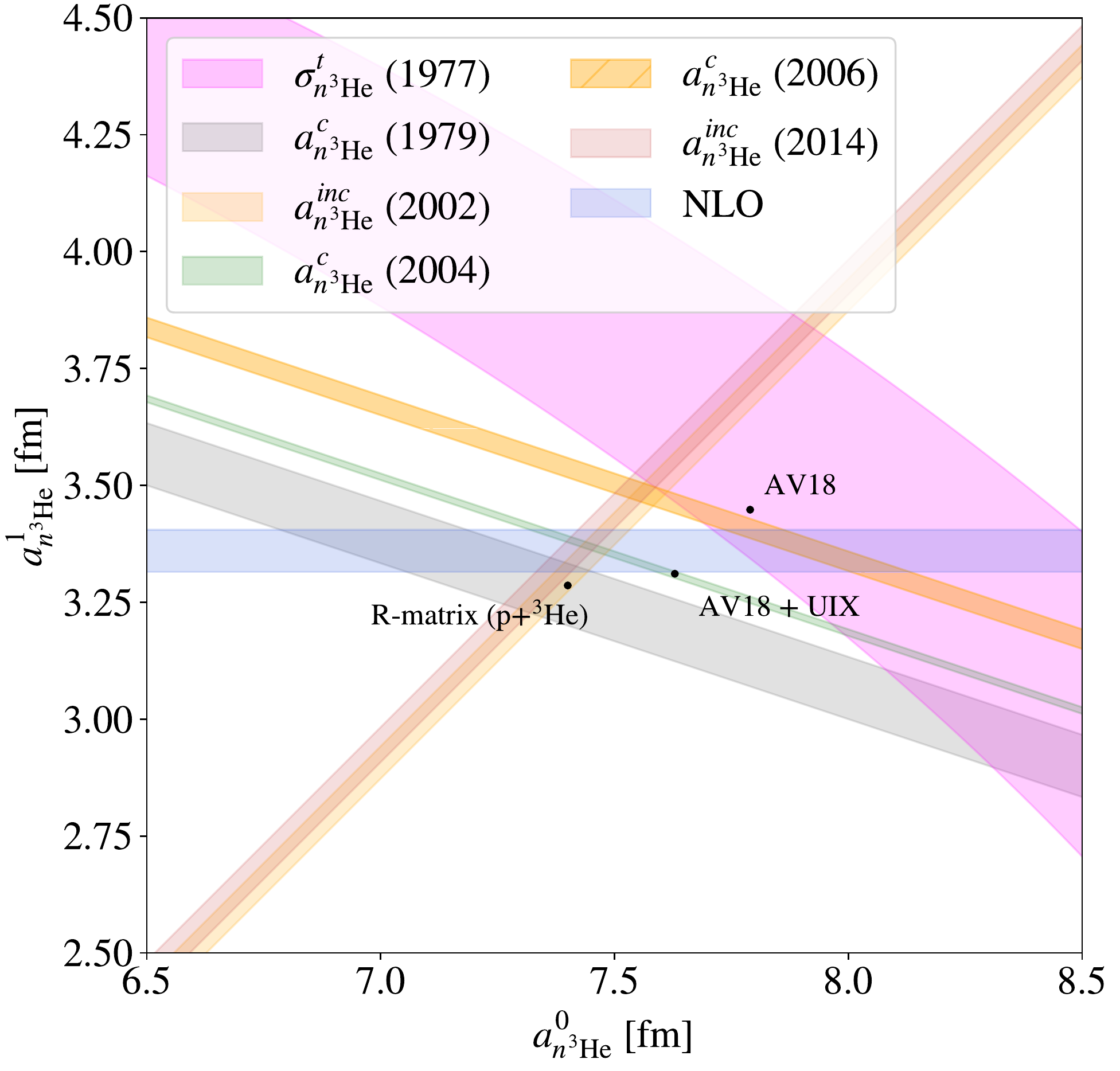}\\
\end{tabular}
\caption{\label{figN3NALL}
Left panel: The $(a^{0}_{nt}; a^{1}_{nt})$ plane with experimental constraints imposed by different measurements -- size of coherent scattering length $a_{nt}^c$ (1972) \cite{DBO72}, (1981) \cite{HRCK81}, (1985) \cite{RTWW85} and total cross-section at zero momentum $\sigma^t_{nt}$ (1972) \cite{DBO72}, $\sigma^t_{nt}$ (1980) \cite{PBS80}. Our \nopieft LO and NLO predictions are shown by red and blue error bars, respectively. Right panel : The $(a^{0}_{n{\rm ^3 He}}; a^{1}_{n{\rm ^3 He}})$ plane with experimental constraints imposed by different measurements -- size of coherent scattering length $a_{n{\rm ^3 He}}^c$ (1979) \cite{kaiser79}, (2004) \cite{huffman04}, (2006) \cite{kettler06}; size of incoherent scattering length $a_{n{\rm ^3 He}}^{\rm inc}$ (2002) \cite{zimmer02}, (2014) \cite{huber09}; and total cross-section at zero momentum $\sigma^t_{n{\rm ^3 He}}$ (1977) \cite{alfimenkov77}. Our $a^{1}_{n{\rm ^3 He}}$ NLO prediction is represented by a blue band. For references to individual theoretical predictions in both panels (black dots/errorbars) see the text.}
\end{figure*}

\subsubsection{Zero momentum $n t$ and $n{\rm ^3He}$ scattering}
Unlike in the $nd$ case, more unsettled situation persists in zero-momentum $n t$ and $n{\rm ^3He}$ scattering. Here, spin-singlet ($a^0_{N3N}$) and spin-triplet ($a^1_{N3N}$) scattering lengths, $N3N \in \left\{ nt , n{\rm ^3He}\right\}$, are constrained by experimentally measured coherent scattering length       
\be
a^c_{N3N} = \frac{1}{4} a^0_{N3N} + \frac{3}{4} a^1_{N3N},
\ee
incoherent scattering length
\be
a^{\rm inc}_{N3N} = \frac{\sqrt{3}}{4} \left[ a^1_{N3N} - a^0_{N3N} \right],
\ee
or the total cross-section at zero momentum
\be
\sigma^t_{N3N} = 4\pi \left[\frac{1}{4} \left( a^0_{N3N} \right)^2 + \frac{3}{4} \left( a^1_{N3N} \right)^2 \right].\vspace{5pt}
\ee

In the left panel of Fig.~\ref{figN3NALL} we show $(a^0_{n t} ; a^1_{n t})$ plane with shaded areas representing possible sets of scattering length values suggested by experimentally measured $a^c_{n t}$ \cite{DBO72,HRCK81,RTWW85} or $\sigma^t_{n t}$ \cite{DBO72,PBS80}. As can be seen from the figure experimental results are mostly in mutual disagreement or burdened by a rather large experimental uncertainty. Adding to the figure also theoretical predictions of microscopic four-body calculations using AV18 $NN$ interaction with and without Urbana IX three-body force \cite{VRK98,LC04}, and $\chi$EFT(N$^3$LO) $NN$ interaction with and without three-body $\chi$EFT(N$^2$LO) force \cite{VGKM20} suggests smaller area of possible $(a^0_{n{\rm ^3H}} ; a^1_{n{\rm ^3H}})$. Interestingly, scattering lengths extracted from $R$-matrix analysis of $p{\rm ^3He}$ scattering data \cite{Hal90} seem to be slightly shifted from these theoretical results. 
At LO \nopieft the size of $a^0_{n t}$ and $a^1_{n t}$ scattering lengths was predicted in Ref.~\cite{Kir13}, here, a rather large theoretical uncertainty was assigned via momentum cutoff variation between $2$~and~$8~{\rm fm^{-1}}$. In the same work, the author numerically demonstrated a correlation between the triton binding energy $B_3$ and $(a^0_{nt}$, $a^1_{n{\rm ^3H}})$ where with decreasing $B_3$ the magnitude of scattering lengths grows. This is in agreement with slightly shifted AV18 and chiral results with no three-body force -- disregarding the three-body interaction in these calculations slightly underestimates experimental $B_3$.

We extract $nt$ $s$-wave scattering parameters in spin-singlet and spin-triplet from our $N-3N$ calculations in $S,I=(0,1)$ and $S,I=(1,1)$ four-body channels, respectively. Note that within our approach $B_3$ is used to renormalize a three-body contact interaction, consequently, $S,I=(1/2,1/2)$ three-body bound state ($3N$) is kept consistently fixed at triton experimental binding energy both at LO and NLO. Taking into account numerical errors and residual cutoff variation for $\lambda \in [1.25,10]~{\rm fm^{-1}}$ we obtain scattering length values
\be
\begin{split}
    \text{LO}~~  &:~~ a^0_{n t}=3.42(50)~{\rm fm}~~~~~a^1_{n t}=3.06(42)~{\rm fm}\\[5pt]
    \text{NLO} &:~~ a^0_{n t}=4.035(65)~~{\rm fm}~~~ a^1_{n t}=3.566(47)~{\rm fm}\\
\end{split}
\nonumber
\ee
which are displayed by red (LO) and blue (NLO) error bars in the left panel of Fig.~\ref{figN3NALL}. Our LO predictions are in agreement with Ref.~\cite{Kir13}. A slightly larger uncertainty of our LO results is induced by a larger $\lambda$ interval considered for theoretical error assessment than in Ref.~\cite{Kir13}. This error estimate is significantly reduced including subleading effective range corrections. At NLO resulting $a^0_{n t}$ and $a^1_{n t}$ are predicted in remarkable agreement with AV18 and $\chi$EFT microscopic results with three-body forces included.     

\begin{table*}
\begin{tabular}{cllcl}
     \hline \hline
     ~~~~~~~~Scattering~~~~~~~~&\multicolumn{2}{c}{Values~~~~~~~~~~~~~}&&~~~~~~~~~~~~~~~~Exp./Model~~~~~~~~\\[5pt] \hline
     $nd$&$a^{1/2}_{nd}$~~~~~~~~~~~~~~~~&$a^{3/2}_{nd}$~~~~~~~~~~~~~~~~&~~~~~~~~~~~~~~~~&\\[5pt]
                     &0.29&5.6&&exp. ($nd$ phase shifts) \cite{OerSea67}\\
                     &0.65(4)&6.35(2)&&exp. (extracted from $\sigma^{t}_{nd}$ and $a^c_{nd}$) \cite{DilKoeNis71}\\
                     &0.645(3)(7)&~~---&&exp. $a^c_{nd}$ + collection of theor. $a^{3/2}_{nd}$ results \cite{schoen03}\\
                     &1.304&6.346& &AV18 \cite{WNKGGS03}\\
                     &0.636&6.437& &AV18 + UIX \cite{WNKGGS03}\\
                     &~~---&6.33(10)&&\nopieft(NLO); nonperturbative \cite{BedKol98}\\
                     &~~---&6.7(7)&&\nopieft(NLO); nonperturbative \cite{BedGri00}\\
                     &~~---&6.354(20)&&\nopieft(N$^2$LO); partial resummation \cite{Gri04}\\
                     &~~---&6.19(30)&&\nopieft(N$^2$LO); perturbative \cite{Van13}\\
                     &0.92(29)            &6.330(14) & &This work \nopieft(NLO); perturbative\\
                     &&&&\\
                     & $r^{1/2}_{nd}$& $r^{3/2}_{nd}$& &\\[5pt] 
                     &1.7&---&&exp. ($nd$ phase shifts) \cite{OerSea67}\\
                     &~~---&1.8(1)&&\nopieft(N$^2$LO); partial resummation \cite{Gri04}\\
                     & 1.74(33)&1.891(91)& &This work \nopieft(NLO); perturbative\\
                     &&&&\\
                     & \multicolumn{2}{c}{$E^{1/2}_{nd}$ (virt.)} && \\ 
                     & \multicolumn{2}{c}{-0.48}& &separable potential \cite{AFT82} \\
                     & \multicolumn{2}{c}{-0.574}& &\nopieft(LO) \cite{GVHK19}\\
                     & \multicolumn{2}{c}{-0.503(31)}& &This work \nopieft(NLO); perturbative\\\hline
    $nt$     & $a^{0}_{nt}$& $a^{1}_{nt}$& &\\[5pt]
                     &3.60(32)&3.10(23)& &\nopieft(LO) \cite{Kir13}\\
                     &4.453&3.325& &$R$-matrix ($p + {\rm ^3He}$) \cite{Hal90}\\
                     &4.28&3.73& &AV18 \cite{VRK98}\\
                     &4.05(10)&3.595(5)& &AV18 + UIX \cite{VRK98,LC04}\\
                     &4.171(63)&3.646(23)& &$\chi$EFT(N$^3$LO) (500 \& 600) \cite{VGKM20}\\
                     &4.046(81)&3.533(26)& &$\chi$EFT($\rm N^3LO$) + $V_3$ $\rm N^2LO$ (500 \& 600) \cite{VGKM20}\\
                     &4.035(65)&3.566(47)&&This work \nopieft(NLO); perturbative\\
                     &&&&\\
                     & $r^{0}_{nt}$& $r^{1}_{nt}$& &\\[5pt]
                     &2.08&1.72& &AV18 + UIX \cite{privCommLaz}\\
                     &2.117(10)&1.743(10)& &$\chi$EFT(N$^3$LO) (500 \& 600) \cite{privCommViv}\\
                     &2.058(4)&1.709(6)& &$\chi$EFT($\rm N^3LO$) + $V_3$ $\rm N^2LO$ (500 \& 600) \cite{privCommViv}\\
                     &2.17(15)&1.76(41)& &This work \nopieft(NLO); perturbative\\\hline
    $n{\rm ^3He}$    & $a^{0}_{n{\rm ^3He}}$& $a^{1}_{n{\rm ^3He}}$& &\\[5pt]
                     &7.790&3.448& &AV18 \cite{HH03}\\
                     &7.629&3.311& &AV18 + UIX \cite{HH03}\\
                     &6.98&3.20& &$R$-matrix ($p + {\rm ^3He}$) \cite{HH03}\\
                     &7.5(6)&~~---& &\nopieft(NLO); nonperturbative \cite{KirGriShu10}\\
                     &~~---& 3.360(45)& &This work \nopieft(NLO); perturbative \\
                     &&&&\\
                     & $r^{0}_{n{\rm ^3He}}$& $r^{1}_{n{\rm ^3He}}$ & &\\[5pt] 
                     &---&1.83(28) & &This work \nopieft(NLO); perturbative\\\hline \hline
                 
\end{tabular}
\caption{\label{table:sum}Theoretical and experimental $nd$, $nt$, and $n{\rm ^3He}$ scattering length and effective range values in fm together with energies of $nd~S=1/2,I=1/2$ virtual state $E_{nd}^{1/2}$ in MeV. We note that in some \nopieft works \cite{GabBedGri00,HamMeh01,BedRupGri03,MarSprVan16,Kon17,RupVagHig19} three-body LEC(s) are fitted to reproduce experimental scattering length $a^{1/2}_{nd}=0.65(4)$~fm \cite{DilKoeNis71}.}
\end{table*}

In the right panel of Fig.~\ref{figN3NALL} we depict $(a^0_{n{\rm ^3He}} ; a^1_{n{\rm ^3He}})$ plane where shaded areas denote possible sets of scattering length values suggested by experimental $a^c_{n{\rm ^3He}}$ \cite{kaiser79,huffman04,kettler06}, $a^{\rm inc}_{n{\rm ^3He}}$ \cite{zimmer02,huber09}, and $\sigma^t_{n{\rm ^3He}}$ \cite{alfimenkov77}. Similarly as in the $n{\rm ^3H}$ case some experimental constraints are in mutual disagreement. There are only few available theoretical results -- microscopic calculation using AV18 interaction with and without Urbana IX three-body force and $R$-matrix analysis of $p{\rm ^3He}$ scattering data \cite{HH03}.

In our \nopieft calculations of $n {\rm ^3 He}$ scattering we fix at NLO the ground state energy of $S,I=(1/2,1/2)$ three-body system at experimental $B({\rm ^3He}) = 7.718$~MeV \cite{tritonBE}, hence we simulate the correct position of the corresponding $3+1$ threshold. The spin-triplet scattering length is then calculated isospin-averaging $a^{1}_{n{\rm ^3He}}=\sqrt{1/2 \left(a^{(1,0)}_{N3N}\right)^2+1/2 \left(a^{(1,1)}_{N3N}\right)^2}$ over $S,I=(1,0)$~and~$(1,1)$ four-body channels,
\be
    \text{NLO} :~~ a^1_{n{\rm ^3He}}=3.360(45)~~{\rm fm},
\nonumber
\ee
where the error is estimated in the same manner as in the case of $n t$ scattering. Our result is then depicted in the right panel of Fig.~\ref{figN3NALL} in a form of a blue horizontal band showing consistency with the previous theoretical studies.

Not considering Coulomb interaction introduces certain shortcomings into our description of spin-singlet $n {\rm ^3 He}$ scattering. First, $t$ and $\rm ^3He$ ground state energies are degenerate, which leads to degenerate $p t$ and $n ^3$He thresholds as well. As a result, the position of $0^+_2$ resonance in $^4$He is not correctly reproduced. The same subsequently holds for the low-energy $n{\rm ^3He}$ scattering in $S,I = (0,0)$ four-body channel which is strongly affected by the resonance position. Therefore, we give here only our result for the isospin-triplet $S,I=(0,1)$ component of $a^{0}_{n{\rm ^3He}}$ scattering length
\be
\text{NLO} :~~ a^{(0,1)}_{n{\rm ^3He}} =4.171(82)~~{\rm fm},
\ee
keeping in mind that its dominant $I=0$ contribution has to be determined in future studies with Coulomb interaction included. 

\section{Conclusion}
We performed a thorough analysis of $s$-wave scattering processes in $A \leq 4$ nuclear systems within the first two orders of \nopieft. Using the stochastic variational method we solved the few-body Schr\"{o}dinger equation for nuclear bound states both inside and outside shallow harmonic oscillator trap. This allowed us to extract $s$-wave scattering phase shifts using the Busch formula in all possible $A$-body spin-isospin channels which support partition into two subclusters. 

Our EFT was fitted to six well-established experimental parameters. At LO we fixed the two-body LECs using the experimental $nn$ spin-singlet and $np$ spin-triplet scattering lengths. The three-body LEC was fitted to the triton ground state binding energy. NLO terms were included perturbatively. At this order, LECs of two-body momentum-dependent interaction were constrained by the experimental effective ranges in $nn$ spin-singlet and $np$ spin-triplet channels. In agreement with the results of Ref.~\cite{BazKirKon19},  We found it necessary to include at NLO four-body force to retain cutoff invariant predictions once all four nucleons are allowed to occupy the $s$-shell, i.e. in the $S,I=(0,0)$ four-body channel.
We thus have to include one four-body LEC, which is fitted to the $0^+$ ground state binding energy of $\rm ^4 He$. 

Using this theory, we calculated elastic $nd$, $dd$, $nt$, and $n{\rm ^3He}$ $s$-wave scattering at LO and NLO \nopieft. For a wide momentum cutoff interval $\lambda \in [1,10]~{\rm fm^{-1}}$ we numerically demonstrated that all our predictions converge with increasing cutoff once $\lambda \gg m_\pi$, i.e. when the momentum cutoff is much larger than the breakup scale of the theory. Our results further show that no NLO four-body force is needed in the $S,I=(0,1)$, $(1,0)$, $(1,1)$, and $(2,0)$ spin-isospin four-body channels, while its inclusion in $S,I=(0,0)$ provides cutoff invariant scattering predictions. The remaining $S,I=(0,2)$, $(1,2)$, $(2,2)$, and $(2,1)$ four-body channels were not studied; however, it is reasonable to assume that since in these channels at least two nucleons are Pauli blocked from the $s$-shell no four-body force will be needed as well.      

Despite the simplicity of the \nopieft approach, all our NLO low energy scattering predictions are in remarkable agreement with the available experimental data or results of other interaction models. In particular, predicted $nt$ scattering lengths values basically coincide with results obtained with AV18+UIX potential \cite{VRK98,LC04} or $\chi$EFT(N$^3$LO) interaction with N$^2$LO three-body force \cite{VGKM20}. For clarity we summarize our $nd$, $nt$, and $n{\rm ^3 He}$ scattering length and effective range predictions together with calculated energy of $nd$ $S=1/2,I=1/2$ virtual state and compare with the available theoretical and experimental results in Tab.~\ref{table:sum}.

The apparent shortcoming of our study is a lack of Coulomb interaction. Its inclusion would allow us to compare our predictions with more extensive $pd$, $dd$, $p {\rm ^3H}$, or $p {\rm ^3He}$ scattering data. The Coulomb interaction is in particular important for a correct description of $p ^3$H and $n ^3$He threshold and the position of a narrow $0^+_2$ resonance in $^4$He. In future work 
it would be interesting to study Coulomb inclusion alongside NLO corrections in $A \geq 4$ systems which, to the best of our knowledge, have not been performed so far.

Our \nopieft calculations should be regarded as a stepping stone towards a study of subleading corrections in heavier nuclear systems beyond $s$-shell. While this work is the first of its kind where effective range corrections are introduced perturbatively in a 4-body nuclear system, $A \geq 5$ nuclear systems remain unexplored. One of the main advantages of Hamiltonian formalism, adopted in our calculations, is a rather straightforward extension to $p$-wave scattering such as $n ^3$H or $n ^4$He, strongly affected by the position of $^4$H and $^5$He resonances, or even to heavier nuclei. So far it has been concluded that at LO nuclear systems beyond $s$-shell do not retain bound states in the zero-range contact limit; instead, they break into bound $s$-shell subclusters \cite{DCKG20,SCKM21}. However, as argued in Ref.~\cite{SCKM21}, including subleading corrections might change the outcome of LO calculations and bring these systems to the bound state region. It would be interesting to explore to which extent this might be done considering NLO terms.  

The form of effective range corrections, methods, and procedures outlined in this work are applicable beyond the scope of nuclear systems. In fact, they might be used in an analysis of LQCD data \cite{EBB20,DS21,parreno21,YBSB22} or even in a study of more exotic systems such as hypernuclei \cite{CBG18,HH19,HH20,SBBGM22}, and mesic nuclei \cite{BBFG17}. As might be seen in our work, including NLO terms dramatically improves the predictive power of the theory and the same outcome might be expected in these systems as well.

\section*{ACKNOWLEDGMENT}
We would like to thank Nir Barnea and Dmitry Petrov for useful discussions and communications. The work of MS was supported by the Pazy Foundation and by the Israel Science Foundation grant 1086/21.



\end{document}